\documentclass[twocolumn]{aastex631}

\usepackage{shortbold}
\usepackage{xspace}
\usepackage{booktabs}

\usepackage{amsmath}

\DeclareMathOperator*{\argmin}{arg\,min}
\newcommand{\hinode}{{\it Hinode}/SOT-SP\xspace}
\newcommand{\hmi}{{{\it SDO}/HMI}\xspace}
\newcommand{\aia}{{{\it SDO}/AIA}\xspace}
\newcommand{\mdi}{{{\it SOHO}/MDI}\xspace}
\newcommand{\vectwo}[2]{\left[ \begin{array}{c} {#1} \\ {#2} \\ \end{array} \right]}
\newcommand{\mattwo}[4]{\left[ \begin{array}{cc} {#1} & {#2} \\ {#3} & {#4} \\ \end{array} \right]}
\newcommand{\diag}{\textrm{diag}}
\newcommand{\xcen}{{\tt XCEN}\xspace}
\newcommand{\ycen}{{\tt YCEN}\xspace}
\newcommand{\xscale}{{\tt XSCALE}\xspace}
\newcommand{\yscale}{{\tt YSCALE}\xspace}
\newcommand{\dxcen}{$\Delta${\tt XCEN}\xspace}
\newcommand{\dycen}{$\Delta${\tt YCEN}\xspace}

\newcommand{\change}[1]{{\textcolor{blue}{{\bf #1}}}}
\shorttitle{Large Scale Spatial Cross-Calibration of Hinode/SOT-SP and SDO/HMI} 
\shortauthors{Fouhey et al.}
\graphicspath{{./}{figures/}}

\begin{document}

\title{Large-Scale Spatial Cross-Calibration of \hinode and \hmi}

\author[0000-0001-5028-5161]{David F. Fouhey}
\affiliation{University of Michigan, Department of Electrical Engineering and Computer Science,
Ann Arbor, MI}

\author[0000-0002-6227-0773]{Richard E. L. Higgins}
\affiliation{University of Michigan, Department of Electrical Engineering and Computer Science,
Ann Arbor, MI}

\author[0000-0003-0176-4312]{Spiro K. Antiochos}
\affiliation{University of Michigan, Department of Climate and Space Sciences and Engineering, Ann Arbor, MI}

\author[0000-0003-3571-8728]{Graham Barnes}
\affiliation{NorthWest Research Associates,
Boulder, CO}

\author[0000-0002-6338-0691]{Marc L. DeRosa}
\affiliation{Lockheed Martin Solar and Astrophysics Laboratory, Palo Alto, CA, USA}

\author[0000-0001-9130-7312]{J. Todd Hoeksema}
\affiliation{Stanford University,
Stanford, CA}

\author[0000-0003-0026-931X]{K. D. Leka}
\affiliation{NorthWest Research Associates,
Boulder, CO}

\author[0000-0002-0671-689X]{Yang Liu}
\affiliation{Stanford University,
Stanford, CA}

\author[0000-0003-1522-4632]{Peter W. Schuck}
\affiliation{NASA GSFC,
Silver Spring, MD}

\author[0000-0001-9360-4951]{Tamas I. Gombosi}
\affiliation{University of Michigan,
Department of Climate and Space,
Center for Space Environment Modelling,
Ann Arbor, MI}



\begin{abstract}
We investigate the cross-calibration of the Hinode/SOT-SP and SDO/HMI instrument meta-data, specifically the correspondence of the scaling and pointing information. Accurate calibration of these datasets gives the correspondence needed by inter-instrument studies and learning-based magnetogram systems, and is required for physically-meaningful photospheric magnetic field vectors. We approach the problem by robustly fitting geometric models on correspondences between images from each instrument's pipeline. This technique is common in computer vision, but several critical details are required when using scanning slit spectrograph data like Hinode/SOT-SP. We apply this technique to data spanning a decade of the Hinode mission.  Our results suggest corrections to the published Level 2 Hinode/SOT-SP data. First, an analysis on approximately 2,700 scans suggests that the reported pixel size in Hinode/SOT-SP Level 2 data is incorrect by around 1\%. Second, analysis of over 12,000 scans show that the pointing information is often incorrect by dozens of arcseconds with a strong bias. Regression of these corrections indicates that thermal effects have caused secular and cyclic drift in Hinode/SOT-SP pointing data over its mission. We offer two solutions. First, direct co-alignment with SDO/HMI data via our procedure can improve alignments for many Hinode/SOT-SP scans. Second, since the pointing errors are predictable, simple post-hoc corrections can substantially improve the pointing. We conclude by illustrating the impact of this updated calibration on derived physical data products needed for research and interpretation. Among other things, our results suggest that the pointing errors induce a hemispheric bias in estimates of radial current density.

\end{abstract}



\section{Introduction} \label{sec:intro}

The Sun's photosphere is co-observed by a variety of instruments that provide tradeoffs between design considerations that range from spectral to spatial sampling. Each instrument independently provides a view of the magnetic field, and by combining multiple instruments with varying capabilities, one can capture phenomena from different vantage points, obtain information that is impossible to extract from one instrument alone, and calibrate the instruments. For instance, \cite{dalda2017statistical} use co-aligned data to better understand the magnetograms produced by the pipelines of \hinode and \hmi. Similarly, our previous work, \cite{Higgins2022} shows how to use co-aligned \hinode and \hmi data to produce synthetic magnetograms from \hmi Stokes vectors with characteristics similar to \hinode data. These efforts on using multiple instruments depend critically on precise co-alignment of the data.

In this paper, we demonstrate and analyze the spatial co-alignment of the {\it Solar Optical Telescope-Spectro-Polarimeter} \citep[SOT-SP;][]{tsuneta2008solar} on the {\it Hinode} satellite \cite[]{kosugi2007hinode} with the {\it Helioseismic and Magnetic Imager} \citep[HMI;][]{Schou2012} on the {\it Solar Dynamics Observatory} \citep[SDO;][]{PesnellThompsonChamberlin2012}. Since the focus of our work is magnetograms, we specifically focus on Level 2 \hinode data~\citep{SPLevel2} inverted by MERLIN~\citep{Lites2006} and for \hmi, the {\tt hmi.ME\_720s\_fd10} series inverted by the \hmi pipeline version of VFISV~\citep{borrero2011vfisv,hmi_invert}. 

We obtain the relative scaling and pointing of the two instruments using techniques that are common in computer vision. We approach the problem via a standard computer vision practice of fitting geometric models (see e.g., the books of~\citealt{Hartley04} and ~\cite{szeliski2010computer}) to correspondences automatically extracted from image data. These correspondences come from the point-based features SIFT~\citep{lowe2004} and ORB~\citep{rublee2011orb} that are extracted at distinctive and easily localizable locations in the image. Because the correspondences are extracted by matching texture in small image regions, they are likely contaminated with outliers. However, by fitting models robustly using RANSAC~\citep{Fischler81} to reject outliers, we are able to  characterize accurately the relationship between the instruments.
This general framework of fitting models to correspondences is well-understood and validated in computer vision and has been applied successfully to tasks ranging from automatic panorama construction~\citep{Brown03} to building precise 3D reconstructions from photo collections~\citep{Snavely07,schoenberger2016sfm} to performing simultaneous localization and mapping~\citep{mur2017orb}, and far more. However, several key adjustments to this standard procedure are needed for the \hinode data because the data comes from a scanning slit spectrograph. This spectrograph captures the Sun at different times of evolution and does not uniformly sample the scan spatially. 
Both challenges can be overcome by modifications to this alignment procedure that we describe in Section \ref{sec:method}.

We apply this technique to thousands of \hinode scans that span nearly a decade and come from both of \hinode's fast and normal modes~\citep{tsuneta2008solar,lites2013hinode}, which acquire data with different effective pixel sizes (described in further detail in Section~\ref{sec:data}). By aligning the data, we can extract parameters that ought to not vary between scans, namely the relative scaling of the instruments, as well as parameters that will vary between scans, namely the relative pointing information and instrument rotation. We can align the data while letting the relative instrument scaling vary freely. By performing this procedure across multiple scans, we can obtain an updated empirical estimate of the relative scaling of the instruments. Once a new scale has been fit, we can estimate updated pointing information and compare with the reported pointing information.  

The first key result of this work, described in Section~\ref{sec:scale} is a suggested correction to the reported \hinode Level 2 data pixel size. Fit scale parameters  suggest that data are consistently explained better with different scaling parameters than is reported. This scale correction is nearly the same for both fast and normal scans and is near constant across observations spanning nearly a decade of \hinode's missions. The fit scale additionally shows no strong correlation with spatial location or time (represented at multiple frequencies). 

The second key result is a suggested update to \hinode's pointing information, described in Section~\ref{sec:pointing}. This information is known to be inaccurate~\cite[Appendix C]{hinodeDataAnalysisGuide} and our approach provides highly accurate correspondence between the scans, providing a way to update the pointing information reported with the \hinode scan. These pointing updates are typically on the order of dozens of arcseconds and show a strong directional bias. Examination of the pointing updates as a function of time and temperature readings suggest that these residuals can be well-explained by secular trends in pointing information during the Hinode mission as well as yearly thermal changes during eclipse season. Co-alignment with \mdi~\citep{scherrer1995solar} further suggests that the pointing error was substantially smaller earlier in the mission. New \hinode data can be, of course, co-registered to new \hmi data to provide improved pointing information. If this registration fails or \hmi is no longer available, our results show that the drift is relatively predictable and one can obtain substantially more accurate pointing information (about an order of magnitude smaller error) via a simple predictive model.

Section~\ref{sec:implications} describes the physical implications of the updated pointing information. The change in pointing information affects both the heliographic coordinates of each pixel as well as distance between the parts of the sun depicted by each pixel. These updates lead to distinctive spatial patterns and, due to the bias in the pointing update, hemispheric bias in quantities such as the total electric current. A summary discussion and concluding remarks are presented in Section~\ref{sec:discussion}.

\section{Data}
\label{sec:data}

We work with two sources of data, namely the pipelines of \hinode and \hmi, focusing on 11 years of \hinode Level 2 data from 1 January 2011 until 31 December 2021 as well as \hmi data that was observed at the same time. The goal of this section is to reintroduce salient aspects of the data acquisition pipelines that are needed to understand the technical conclusions of the paper. We begin by reprising the data pipeline, consisting of an overview of the acquisition process (Section~\ref{sec:data_overview}) and a description of how scale and pointing information changes in the \hinode and \hmi data pipelines (Section~\ref{sec:data_scalevariance}. A more full description of the characteristics of the data products is beyond the scope of the paper. We refer the interested reader to~\cite{tsuneta2008solar} and \cite{Hoeksema2014} for descriptions of the data products of \hinode and \hmi respectively. We then discuss the specific data used in this paper. First, we introduce the data filtering process  (Section~\ref{sec:data_scanid}) and then the datasets that we use (Section~\ref{sec:data_dataset}).

\subsection{Overview of Data Acquisition Pipelines}
\label{sec:data_overview}

\hmi is an imager that prioritizes the acquisition of {\it spatially} coherent observations. While we show images with multiple channels from \hmi, its underlying observation is a single $4096 \times 4096$ full-disk image of the full sun. Over time, multiple such images with different passbands and polarization states are acquired and analyzed in order to recover the Stokes vector. The Stokes vector observation at each pixel is then inverted to produce a magnetogram that is recorded at a nominal time. 

\hinode, on the other hand, is a scanning-slit spectrograph that prioritizes spectrally coherent observations. While we show data from  \hinode throughout the paper as images, its underlying observation is a single vertical scanline with dense spectral sampling of the Stokes vector. These scanlines form the basis of \hinode's Level 1 data. In the data pipeline, a number of these scanlines are concatenated horizontally and the Stokes vector at each pixel in each scanline is inverted. By placing $W$ scanlines side-by-side, this produces a $H \times W$ magnetogram that is reported as the \hinode Level 2 data~\citep{SPLevel2}. For the data used in this study, the acquisition of all the scanlines in a magnetogram takes roughly 30 minutes to an hour.  These magnetograms are not acquired with a regular cadence, since \hinode has targets of interest selected and then acquired. Given the difference in data acquisition times, for any given \hinode scan, there are likely multiple contemporaneous \hmi scans. 

The \hinode acquisition process means that the resulting magnetogram is not at all like \hmi's image: each x coordinate in a \hinode magnetogram represents a scan {\it index}, rather than a scan {\it location}. During acquisition, the spectrograph makes observations at slits, with its position controlled by a tracker. Except in rare cases where the observed slit positions are perfectly evenly spaced, the pattern of the observations is {\it not} uniform but instead contains jumps where the data from individual slit positions are corrupt or otherwise unusable. Since the position of these jumps is known and recorded in the {\tt Mechanical\_Slit\_Position} field, the $H \times W$ magnetogram can instead be thought of as shorthand for a wider, $H \times W'$ image (with $W' \ge W$) that has columns missing. While the wider $H \times W'$ image has missing data, it {\it can} be treated as an image where the $x$ coordinate represents a location just like the $y$ coordinate.

Throughout the paper, we use the word {\it scan}. In the case of \hmi, this refers to a single vector magnetogram from \hmi at a given nominal timestamp. For \hinode, this refers to an accumulation of scanlines into a single $H \times W$ vector magnetogram, also with a nominal timestamp. We stress, however, that the magnetogram cannot be treated as an image and that throughout, we use the information about the slit positions to account for non-uniform scans. In the code release for the paper (to be provided before publication), we will provide demonstration code that does this.

\subsection{Scales and Pointing Throughout the \hinode and \hmi Pipelines}
\label{sec:data_scalevariance}

Due to the flexibility of the \hinode instrument, the pixels of its magnetograms come in a variety of scales. We focus on two types that are roughly square, namely {\it normal} mode and {\it fast} mode. In normal mode, the \hinode Level 2 data headers report a pixel size of 0.15999\arcsec~for \yscale (the size of the pixel in y) and 0.1486\arcsec~for \xscale (the size of the pixel in x). Fast mode scans have double this pixel size due to the ability of the instrument to combine data from neighboring pixels prior to their transmission back to Earth, and due to their smaller telemetry requirements, fast scans are far more common than normal scans in the date range we study. There are other \hinode scans whose pixels have even more non-square aspect ratios; we ignore these in our study because they are substantially less frequent.

While the \hinode Level 2 data reports pixel sizes of 0.15999\arcsec~and 0.1486\arcsec, this is not the case throughout the pipeline. The Level 1 data reports that {\tt CDELT2} (equivalent to \yscale) is 0.1585\arcsec~and the nominal scanning step size (roughly equivalent to \xscale) is 0.1476\arcsec. The reason for the change is an earlier cross-instrument spatial calibration study from~\cite{centeno2009hinode} that calibrated a continuum image synthesized from \hinode and the G band from {\it Hinode}/BFI. Our results contradict this change and suggest that \hinode's Level 1's reported scale is likely correct. We discuss this contradiction in more detail in Section~\ref{sec:discussion}. 

Because \hinode does not observe the full disk, its pointing information is obtained via the {\it Hinode}'s attitude and orbit control system (AOCS)~\citep{kosugi2007hinode}. The AOCS obtains its pointing information from sun sensors and star trackers. This information is used directly to calculate \xcen and \ycen via the {\tt SC\_ATTX} and {\tt SC\_ATTY} keywords and calibration constants. As discussed in \cite{hinodeDataAnalysisGuide}, this pointing information is known to be inaccurate.

In comparison, \hmi is substantially easier to calibrate because it can observe the full disk of the Sun. Its estimated {\tt CDELT1/2} varies slightly over the period that we study with
values ranging from
0.50400\arcsec~to 0.50439\arcsec. For simplicity, throughout the analysis we use a nominal scale of 0.50428\arcsec, the median reported {\tt CDELT1/2} of the HMI scans used in this analysis. This variance has no consequence on the results since the range of scales amounts to a variation of 0.076\% of the median value. An earlier version of this study used a nominal {\tt CDELT1/2} closer to the bottom range of \hmi's scale and produced numbers that were generally the same to three significant figures.

\subsection{\hinode Scan Type Identification}
\label{sec:data_scanid}

The goal of our analysis is to investigate the co-alignment between \hinode and \hmi, and so we automatically identify a subset of the \hinode scans where this is most likely to succeed. We select scans that: (1) have near-square pixel aspect ratio; (2) have monotonic and relatively smooth spectrograph slit position during capture; and (3) are likely on disk as opposed to showing the poles. We next describe how we operationalize these definitions. 

We work with normal (\xscale = 0.1486\arcsec , \yscale = 0.1599\arcsec) and fast (\xscale = 0.297\arcsec, \yscale = 0.320\arcsec) scans, which we identify by comparing reported \xscale and \yscale with fixed values.

We select for scans that are sufficiently smooth to function closely to images. While we introduce techniques that handle non-uniform sampling due to jumps in slit position, scans that substantially deviate from a uniform scanning pattern are likely to introduce noise. Therefore we reject scans based on two criteria using the {\tt Mechanical\_Slit\_Position} data. First, if slit position does not monotonically increase, we reject the scan. The slit position may not monotonically increase if, for instance, \hinode repeatedly scans an active region in a looping pattern. These are insufficiently frequent to merit a special procedure to further split these into images. Second, we compute the increase in slit position from index to index and compare it to a typical increase. Slit position increases that are $10\times$ or more than the median slit increase are identified as discontinuities, and we reject scans with a discontinuity in the middle 96\% of the slit. These have large jumps in slit position, and our unreliable for registration.

Finally, we automatically identify polar scans by comparing the top and bottom rows of the image with the median continuum intensity. We remove polar scans since they tend to have too few features to be reliably aligned with our proposed methods. If more than the majority of pixels in five or more rows have continuum intensity (as recorded by {\tt Original\_Continuum\_Intensity}) less than half the median continuum intensity value, we identify the image as polar. 

\subsection{Datasets}
\label{sec:data_dataset}

We use a large dataset of \hinode scans as well as two subsets of this dataset. Pilot studies with smaller subsets of data identified that pointing was highly varied but scale was more or less constant. Given that pointing was highly varied, we used the full dataset in order to get a full picture of the pointing information. On the other hand, since scale did not vary over time and space and the validation done was computationally expensive, we present analysis of scale only on the subset.

Our full dataset, which we refer to as the {\it Full Datset}, is all the scans from 1 January 2011 through 31 December 2021. This dataset comprises 23,430 scans, of which 16,564 meet our criterion, and 12,062 can be co-aligned using our methods. These 12,062 scans are aligned with 50,943 \hmi observations (amounting to over 40TB of \hmi data). 
We use two large random subsets of the main dataset for analyzing the relative scale of the two instruments. Our {\it Fast Subset} was obtained by: selecting a random 1500 fast scan subset of the full dataset, of which 988 met our criterion, and of which 760 scans can be co-aligned with our methods. Our {\it Normal Subset} was obtained by taking all the normal scans in the dataset that met our criterion, of which 503 scans can be co-aligned with our methods.

\section{Alignment}
\label{sec:method}

\begin{figure}[t]
    \centering
    \includegraphics[width=\linewidth]{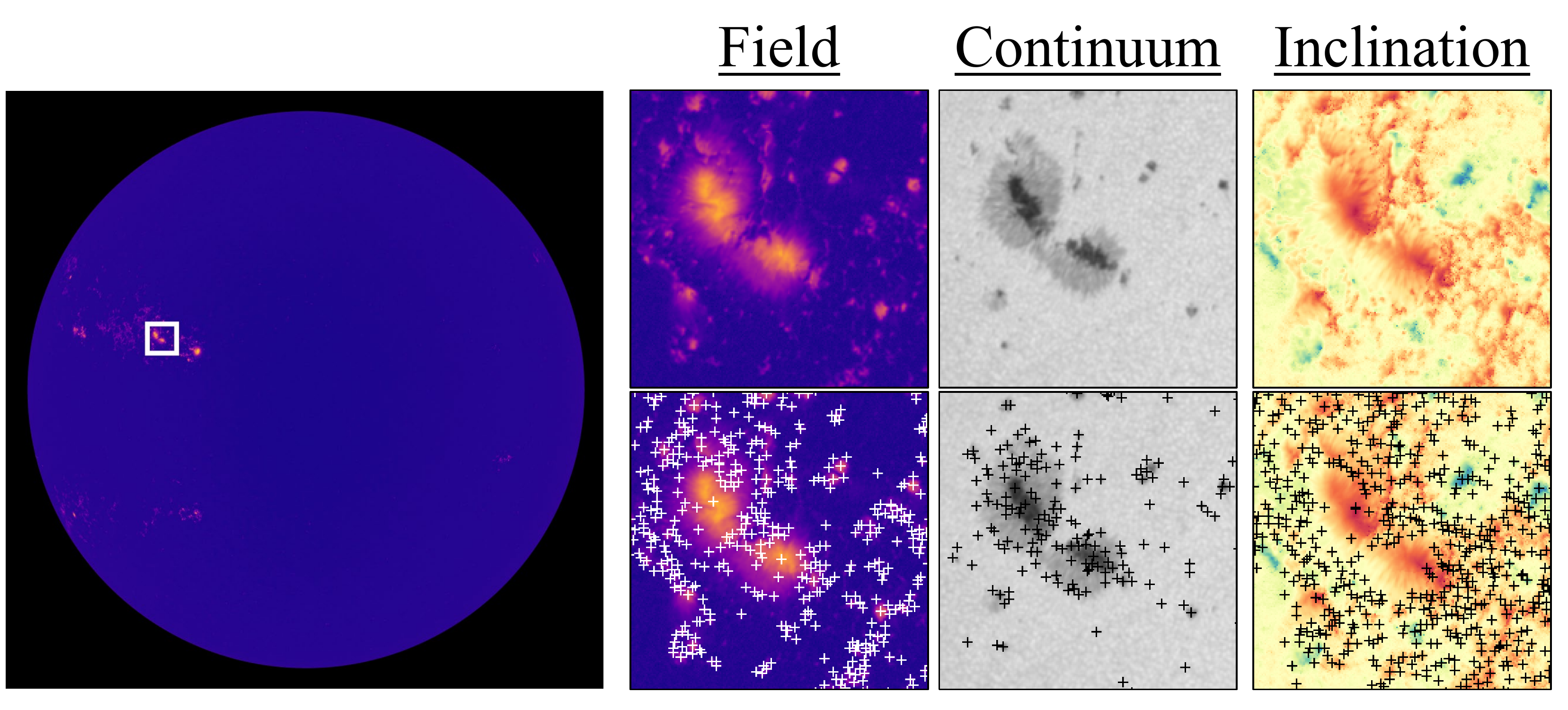}
    \caption{{\bf Locations of SIFT features found on \hmi data.} ORB features are found at similar distinctive regions in the image, but slightly different due to differences in the precise detection scheme In both cases, the features can then be matched to features found on \hinode data to generate correspondences between the two datasets. These correspondences contain both outliers and noise, but the fitting methods used are robust to both.}
    \label{fig:siftlocs}
\end{figure}

\begin{figure*}[t]
    \centering
    \includegraphics[width=\linewidth]{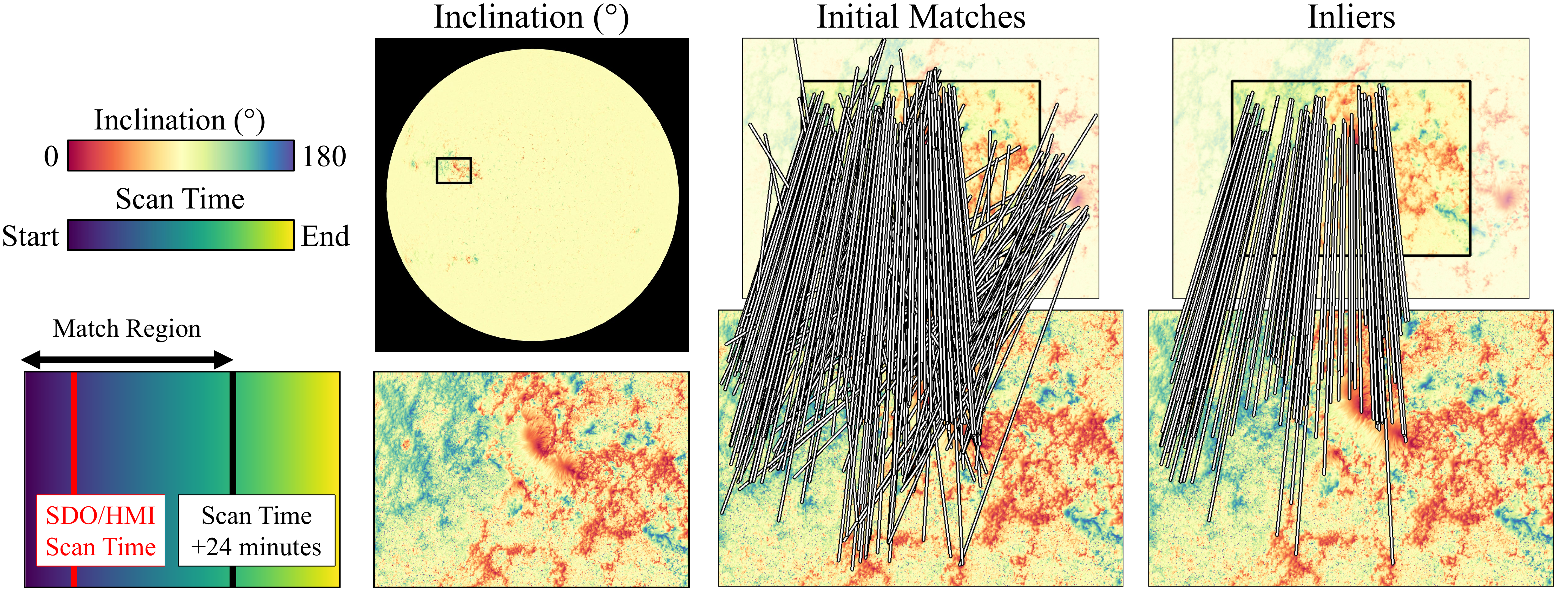}
    \caption{{\bf Diagram of Matching and Transformation Fitting.} Given a Level 2 \hinode scan and \hmi scan, one can identify the contemporaneous part of the \hinode scan and filter to the correspondences on contemporaneous \hinode pixels. For clarity, we show only a subset of the matches that originate near the final fit. These have many outliers. After robust fitting, there are a far smaller number of inliers that are well-described by a single model. For clarity, the diagram shows the Level 2 \hinode scan as it is stored, a set of columns next to each other. In practice, each correspondence's x coordinate is mapped to the slit position to account for the possibility of non-uniform spacing. }
    \label{fig:matches}
\end{figure*}

Our procedure seeks to find a parametric model with a small number of parameters that co-aligns features automatically found in and matched between \hinode and \hmi data. Some of the model parameters, such as the relative scaling of each camera's pixels, do not change between scans. Others, such as the relative translation and rotation, change between the scans. These transformations are fit on point correspondences with a combination of non-linear least-squares~\citep{marquardt1963algorithm} and a robust fitting technique named RANSAC~\citep{Fischler81}. Once fit, one can examine the parameters that were fit to data to both estimate the relative scaling of the instruments' observations as well as estimate the pointing information.

\subsection{Features and Feature Correspondence}
\label{sec:features}

We fit the models on SIFT \citep{lowe2004} and ORB \citep{rublee2011orb} features extracted on similar data from both instruments. These features are commonly used in photogrammetry applications for high-precision measurements, however a full description is beyond the scope of the work. Briefly, given two images $\IB$ and $\IB'$, each correspondence pipeline can generate correspondences between two images $[x,y] \leftrightarrow [x',y']$ such
that point $[x,y]$ in image $\IB$ and point $[x',y']$ in image $\IB'$ both depict the same phenomena. Colloquially, image $\IB$ at $[x,y]$ is said to ``match'' image $\IB'$ at $[x',y']$. These correspondences are based on local intensity information in the image, extracted at distinctive parts of the image (often found via an approximation of the Laplacian of Gaussian filter). However, since the features are matched based on local intensity, there are many {\it outliers}, or  correspondences between locations that look similar but are not the same. 

One challenge is that the \hinode is a scanning slit spectrograph that takes tens of minutes to finish a spatial scan. During this time, multiple complete sets of spectral- and polarimetric images are acquired by \hmi.
Given a set of correspondences $[x_i, y_i] \leftrightarrow [x'_i, y'_i]$ between a single \hinode scan and single \hmi scan taken at time $t_\textrm{hmi}$, one can further filter correspondences based on time. Each $x$ coordinate in a \hinode scan is associated with a timestamp of the scan $t(x)$ given by the {\tt Times} field. One can simply remove correspondences $[x_i, y_i] \leftrightarrow [x'_i, y'_i]$ where $|t(x_i) - t_\textrm{hmi}| > \delta$ where $\delta$ is an empirically tunable parameter set. We set $\delta$ to 24 minutes to strike a balance between two competing goals. If $\delta$ is too small, the correspondences used for transformation will come from a small time window and therefore come from a small number of scanlines. Estimating the scale in $x$ with a few scanlines will be unreliable since the noise in localization of the points of interest will be a substantial fraction of the range of the data. On the other hand, while making $\delta$ bigger ensures that localization noise
is a small fraction of the range of the data, making $\delta$ too big will result in correspondences coming from substantially different observations.  

Practically speaking, this means that given a \hinode scan and a \hmi scan from nearly the same time, one can automatically generate correspondences between \hmi data and parts of the \hinode scan that were taken within a time window near the \hmi scan.

\subsection{Data and Data Preprocessing}

One strong advantage of the correspondence-based approach is that the quantities that are matched do not have to be precisely the same. In fact, the methods used were designed to be robust to substantial non-linear illumination changes as well as changes in viewing angle~\citep{mikolajczyk2005performance}. Mathematically, this stems from the fact that SIFT features are based on gradient orientation angles that have been spatially-histogrammed. The use of orientation {\it angle} rather than orientation {\it value} gives invariance to the scaling of the quantities, and the use of a spatial histogram gives invariance to small pixel perturbations. Similarly, ORB features are based on relative intensity comparisons after smoothing, which give similar invariance to scaling and perturbations. 

This flexiblity to extract correspondences that are related (e.g., by an unknown nonlinear transformation) but not identical is appealing for cross-instrument analysis. Some quantities that are good for generating correspondence refer to different quantities: the \hinode field represents field strength while the field reported by \hmi is the flux density with an assumed magnetic fill factor of unity. Even when data refer to the same physical quantity, there can be differing characteristics of the pipeline outputs: as \cite{dalda2017statistical} and \cite{Higgins2022} illustrate,  \hmi and  \hinode inclination angles tend to be farther from the imaging plane ($90^\circ$) in plage regions compared to \hmi.  These differences arise from differing instrument characteristics (e.g., spectral sampling) as well as treatment of key Milne-Eddington model parameters (e.g., whether a magnetic fill fraction $\alpha$ is fit as in \hinode or assumed as \hmi)~\citep{btrans_bias_2}.

We generate correspondences on multiple pairs of data from \hinode Level 2 and \hmi ~{\tt hmi.ME\_720s\_fd10} data. The first pair of data we match is the reported field strengths of the pipelines ({\tt Field\_Strength} for \hinode and {\tt field} for \hmi). Note again, that \hmi's field is flux density with an assumed unity fill fraction. The second and third are inclination, ({\tt Field\_Inclination} in \hinode and {\tt inclination} in \hmi) and azimuth ({\tt Field\_Azimuth} in \hinode and {\tt azimuth} in \hmi), after flipping the HMI data to acocunt for differences in {\tt CROTA2}. Finally, we generate correspondences from continuum intensity ({\tt Original Continuum Intensity} and {\tt continuum} from {\tt hmi.ic\_720s}); polarization and field strength ({\tt Polarization} and {\tt field}); and finally Stokes V/circular polarization, where \hinode's {\tt StokesV\_Magnitude} is compared with each of the six V filtergrams from \hmi's {\tt hmi.s\_720s} series. 

The feature methods that we use (SIFT, ORB) were designed for consumer image formats, and so operate on grayscale unsigned 8-bit images. We linearly remap the data to {\tt uint8} by mapping each value $x$ to $255(x-v_\textrm{min})/(v_\textrm{max}-v_\textrm{min})$ and clipping to between 0 and 255. We set the linear range with the goal of ensuring the 255 values are used on the bulk of the distribution, and set $v_\textrm{min}$ and $v_{\textrm{max}}$ to: 0 and 2000 for field strength (to ensure that plage has sufficient contrast); and 20 and 160 for inclination (to avoid spending a large fraction of the range on rare values); and the 1st and 99th percentile of the data (excluding NaNs) otherwise (again, to avoid using much of the range on rare values).

\subsection{Fitting Image Transformation Models}
\label{sec:fitting}

We fit a series of transformations that describe the relationship between the two pieces of data. Once fit, each transformation can be expressed as an affine transformation from \hinode pixel coordiantes to \hmi pixel coordinates, or $f([x,y]) = [x', y'] = \AB [x,y] + [t_x, t_y]$ although each model puts constraints on the form of $\AB$. This affine transformation can, of course, be inverted to enable inverse warping, i.e., to enable estimating the \hinode values at each \hmi pixel. 

While one could identify the translation between the scans using cross-correlation, fitting a transformation on a set of point correspondences offers substantial advantages over cross-correlation. The most important advantage is that correspondences enable the rapid and fine-grained exploration of parameters like scaling and rotation. For instance, to identify whether a scaling of $98.2\%$ or $98.3\%$ better explains the data, one simply computes an alignment error between a set of a few thousand correspondences. This error can then be minimized by a standard non-linear least-squares method. In contrast, using cross-correlation entails resampling the template image (e.g., the \hinode level 2 scan) and performing a cross-correlation. As an additional benefit, one can integrate multiple potential sources of information without having to account for differences in the scaling of the quantities; instead, one just concatenates correspondences from each data source.

Our full assumed transformation takes the form of a scaling, rotation, and translation, or
\begin{equation}
\label{eqn:full}
f([x,y]) = \RB(\theta,c_x,c_y) \mattwo{s_x}{0}{0}{s_y} \vectwo{x}{y} + \vectwo{t_x}{t_y},
\end{equation}
where $\RB(\theta,c_x,c_y)$ is a rotation by $\theta$ about a point $c_x, c_y$ (assumed to be the projection of $0^\circ$,$0^\circ$)  Corresponding points $[x,y] \leftrightarrow [x',y']$ are given, and the remaining parameters are either fixed to presumed values (e.g., $s_x$ can be obtained by the relative scaling of the two sensors) or fit to data
to minimize the squared Euclidean norm of their difference, or $||f([x,y]) - [x',y']||_2^2$. In particular, given a set of correspondences and any fixed parameters, one can solve for the free parameters that best ensure equation~(\ref{eqn:full}) matches each $[x',y']$ in the least-squares sense via nonlinear optimization to minimize the sum of squared distances between the correspondence location (i.e., $[x',y']$ and modeled correspondence location (i.e.,
$\RB\,\diag([s_x, s_y]) [x,y] + [t_x, t_y]$). Throughout, we fit transformations using Levenberg-Marquardt \citep{levenberg1944method,marquardt1963algorithm}.

\begin{figure}
    \centering
    \includegraphics[width=\linewidth]{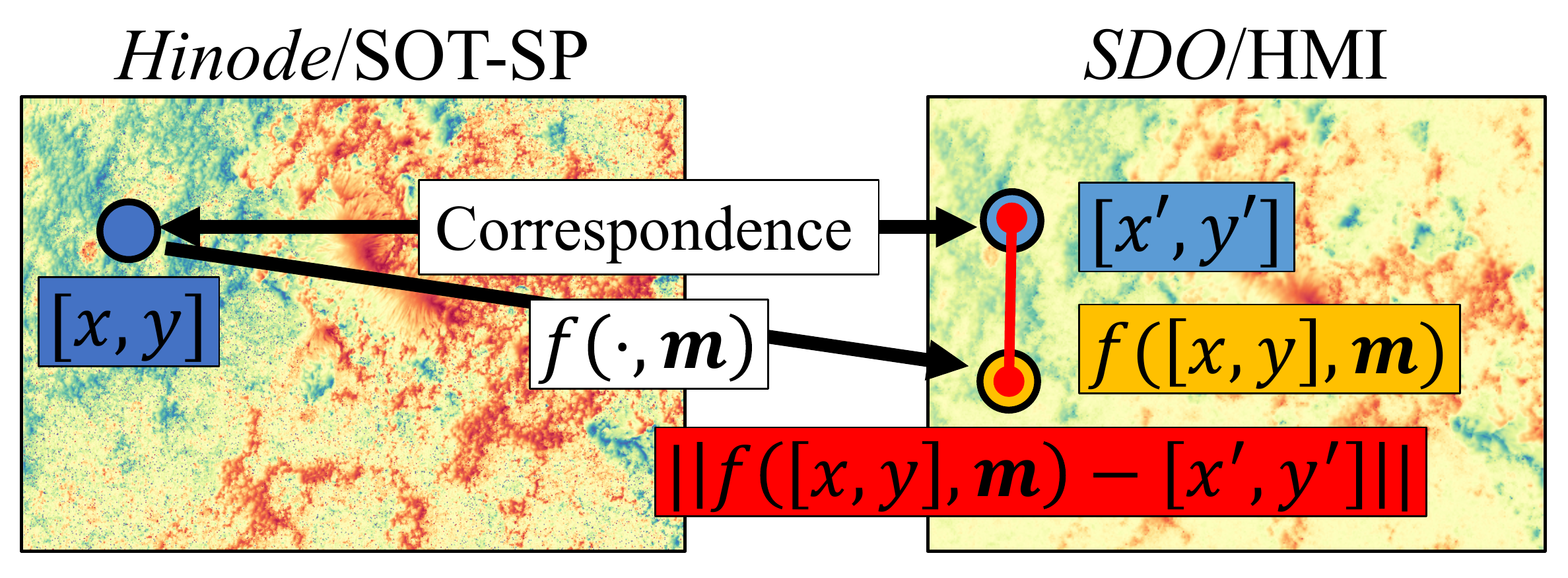}
    \caption{{\bf The matching setup.} Given a correspondence $[x,y] \leftrightarrow [x',y']$, we aim to find a transformation $f$, parameterized by $\mB$, that maps $[x,y]$ to closely match $[x',y']$. Without outliers (i.e., correspondences that are drawn at random), one can safely minimize the squared distance $||[x',y']-f([x,y],\mB)||_2^2$. With outliers, one instead maximizes the number of correspondences satisfying $||[x',y']-f([x,y],\mB)||_2 < \epsilon$.}
    \label{fig:my_label}
\end{figure}

The model expressed in equation~(\ref{eqn:full}) fits five parameters: rotation angle $\theta$, anisotropic scaling $s_x$, $s_y$, and translation $t_x$, $t_y$. In theory, the scaling parameters $s_x$ and $s_y$ are fixed parameters that do not change from scan to scan. The other parameters, $\theta$, $t_x$ and $t_y$ are fit per-scan and indicate the relative pointing information. However, by fitting $s_x$ and $s_y$, we can investigate what scale parameters best explain correspondence data. We additionally analyze how well models that assume different $s_x$ and $s_y$ explain the correspondence data. Finally, with $s_x$ and $s_y$ fixed, one can use the $\theta$, $t_x$ and $t_y$ fit to correspondences to estimate pointing information.

For convenience, we also use a simplified transformation consisting of an assumed lack of rotation (i.e., $\theta=0$), an {\it assumed} scaling, and a fit translation. In particular, this transformation assumes that
\begin{equation}
\label{eqn:simple}
f([x,y]) = \RB(0,c_x, c_y) \mattwo{s_x}{0}{0}{s_y} \vectwo{x}{y} + \vectwo{t_x}{t_y},
\end{equation}
where $s_x, s_y$ are fixed parameters and $\RB(0,c_x, c_y) = I$. The only free parameters are a translation vector $[t_x, t_y]$, which can be solved for in closed form with a single correspondence $[x,y] \leftrightarrow [x', y']$. The transformation of equation~(\ref{eqn:simple}) is useful for the fast screening of outliers: its assumption of $\theta = 0$ is fairly close to reality and so it usually describes correct correspondences fairly accurately. Since it can be fit to a single correspondence, one can simply enumerate all correspondences to examine potential translations supported by the data.

For validation, one can also fit an affine transformation, which takes the 
form 
\begin{equation}
\label{eqn:affine}
f(x,y) = \AB \vectwo{x}{y} + \vectwo{t_x}{t_y}
\end{equation}
where the parameters $\AB \in \mathbb{R}^{2\times2}$ with no constraints and $[t_x,t_y]$ are fit to data. The transformation is slightly incorrect since it has an extra degree of freedom: due to the SVD, $\AB$ can be always factored as $\UB\,\diag([\sigma_1, \sigma_2])\,\VB^T$ where $\UB,\VB$ are rotations, and accordingly this transformation is equivalent to a rotation, scaling, and rotation followed by a translation. This extra degree of freedom enables the affine transform to also include a shearing. While the transformation is incorrect, it is useful since it can provide validation of the fits that is independent of any errors with regards to the center of rotation $c_x, c_y$. Moreover, the fit can be solved in closed form since equation~(\ref{eqn:affine}) can be fit directly using ordinary least-squares. 

\subsection{Robustly Fitting Models}
\label{sec:robustfitting}

The correspondences found in Section~\ref{sec:features} are contaminated with large numbers of outliers, which poses a problem for model fitting. In particular the transformation models described in Section~\ref{sec:fitting} are fit using least-squares and are therefore highly sensitive to outliers. We overcome the challenge of outliers via RANSAC~\citep{Fischler81}, a simple but effective mechanism for model-fitting in the presence of the outliers. 

Suppose we are given a set of $N$ correspondences $[x_i, y_i] \leftrightarrow [x'_i, y'_i]$, a model with parameters $\mB$, and a forward modeling function $f$ such that $f([x_i, y_i],\mB)$ should ideally match $[x'_i, y'_i]$. For example for the affine model (Eqn.~\ref{eqn:affine}), the parameter vector $\mB$ is an element of $\mathbb{R}^6$ and encapsulates both $\AB$ and $[t_x,t_y]$. Then $f([x_i,y_i],\mB)$ computes $\AB [x,y] + [t_x, t_y]$. Least squares aims to find a model that minimize the sum of squared errors, i.e., $\sum_{i=1}^N ||[x'_i, y'_i] - f([x_i, y_i],\mB)||_2^2$. 

Instead of trying to explain {\it every} data point well, RANSAC aims to maximize the number of correspondences that are well-explained by the model, or $\sum_{i=1}^{N} ||[x'_i, y'_i] - f([x_i,y_i], \mB)||_2 \le \epsilon$ for some empirically set parameter $\epsilon$. This set of points that are well-explained by the model is referred to as the model's {\it inliers} (in contrast to outliers). RANSAC proceeds by generating and then checking $k$ hypothesized models $\mB_1, \ldots, \mB_k$. For each hypothesized model $\mB_j$, one computes the inliers associated with $\mB_j$ (i.e.,  $\{i: ||[x'_i, y'_i] - f([x_i, y_i], \mB_j)||_2^2 \le \epsilon\}$). The model with the most inliers is considered the ``best model'', and the best model's inliers are assumed to be inliers for the true transformation. Therefore, one usually re-fits the model in a least-squares sense on the best model's inliers, or $\argmin_{\mB} \sum_{i \in \mathcal{I}} ||[x'_i,y'_i] - f([x_i,y_i],\mB)||_2^2$ if $\mathcal{I}$ are the indices of the inliers.

The hypothetical models $m_1, \ldots, m_k$ are generated by fitting models on small sets of correspondences that are selected via exhaustive or random search. RANSAC fits hypothetical models on small numbers of correspondences, which minimizes the chance that an outlier is included and contaminates the fit. For some models, one can directly enumerate the space of potential models. For instance, the simplified model of equation~(\ref{eqn:simple}) has two free parameters (for translation) and can be exactly fit on one correspondence. Thus, one can simply generate a set of hypothetical models by trying the translation fit on each of the $N$ correspondences. In other cases, (e.g., the affine model in equation~\ref{eqn:affine}), the space of possible models is too large: since the affine model has six parameters and each correspondence provides two equations, model fitting requires at least 3 correspondences. With large numbers of correspondences, ${N}\choose{3}$ is too large to search directly, and so one resorts to randomly sampling. Specifically, one samples $k$ triplets of correspondences without replacement, where $k$ is an empirical parameter.

Practically speaking, this means that a RANSAC fit of an affine model consists of: (1) randomly sampling $k$ triplets of points without replacement and fitting an exact model $m_j$ and counting how many correspondences are well-described by $m_j$; (2) taking the inliers of the model with the most inliers, and then performing a least-squares fit of equation~(\ref{eqn:affine}) on these inliers. 

\subsection{Accounting for Non-Uniform Slit Motion}
\label{sec:method_slit}

As described in Section~\ref{sec:data}, the $H \times W$ Level 2 magnetogram produced by \hinode is shorthand for a larger $H \times W'$ magnetogram $(W' \ge W)$ with missing columns representing slit positions that have been skipped. The spectrograph slit position defines a mapping from each column in the original image ${1, \ldots, W}$ (representing a scan index) to another column in the larger image ${1, \ldots, W'}$ (representing a slit position). We use this mapping to interpolate the column indices of the locations in the scan indexed image, enabling a fit to be performed on the data as if the $x$ coordinate referred to a slit position. Thus, a feature for correspondence found at $x$ location pixel $45$ may be relocated to pixel $49$ if there were four missing columns at the start of the scan. The methods used~\citep{lowe2004,rublee2011orb} estimate the locations of regions of interest at sub-pixel locations, which are handled similarly: a location at pixel $45.7$ would be relocated to $49.7$.

The use of the correspondences where $x$ refers to the slit position rather than the scan number requires only two small changes when one transforms data between \hinode level 2 data and \hmi. When one warps \hmi to the \hinode scan index grid, one first must warp to the grid defined by slit positions, and then drop the missing columns. When one warps \hinode data to the \hmi grid, one must first fill in the missing columns. Our later analysis excludes scans with substantial numbers of missing columns, so this linear interpolation is unlikely to cause substantial difficulties.

\subsection{The Final Models and Model-Fits}

These fitting procedures begin with inlier identification via RANSAC, which produces a best model and corresponding best inliers. Inlier identification is followed by least-squares fitting, in which a model is fit to the best inliers via least-squares.

\begin{enumerate}
    \item The full model fits $\theta, s_x, s_y, t_x, t_y$ to the data after identifying inliers with the simplified model, assuming the theoretical Level 2 values of $s_x,s_y$. To find the simplified model, one fits and then tries all hypothetical models; these models are generated by all the correspondences. This simplified model's inliers are then refitted with the full model (i.e., $\theta, s_x, s_y, t_x, t_y$) by nonlinear least-squares; the full model is initialized with the translation vector from the best-fitting model, the theoretical Level 2 values for $s_x, s_y$, and a rotation of 0. 
\item 
The theoretical Level 2 model fits $\theta, t_x, t_y$ and with $s_x, s_y$ fixed at their theoretical values (i.e., the relative scaling of {\tt XSCALE} using the values in the headers). Like the full model, one identifies a set of inliers via the simplified transformation while assuming $\theta = 0$. These inliers are used for a nonlinear least-squares fit for $\theta, t_x, t_y$ that is initialized with $\theta=0$ and the translation set to the best model from RANSAC.

\item 
The empirical model is identical to the theoretical model, except it assumes a given empirical correction to $s_x$ and $s_y$ that we identify in Section~\ref{sec:scale}.

\item
The affine model that fits $\AB, t_x, t_y$. We identify a set of inliers by fitting $100K$ hypothetical affine models on random triplets from the dataset. We then fit an affine model via least-squares on the inliers of the best model. After $\AB, t_x, t_y$ are fit to data, $\AB$ can subsequently be factored via the SVD to give two scales. 
\end{enumerate}

Throughout, our definition of an inlier is a correspondence where the error 
$||f([x,y]-[x',y']||$
is less than $\epsilon=1$ HMI pixel (i.e., ${\approx}0.5\arcsec$). When fitting a model with no rotation for pre-screening inliers, we use a threshold of $\epsilon=2.5$ HMI pixels (i.e., ${\approx}1.25\arcsec$) since this model is inaccurate due to its failure to account for rotation.

\section{Estimating Relative Scale}
\label{sec:scale}

\begin{figure}
\centering
\begin{tabular}{c@{}c}
    \includegraphics[width=0.48\linewidth]{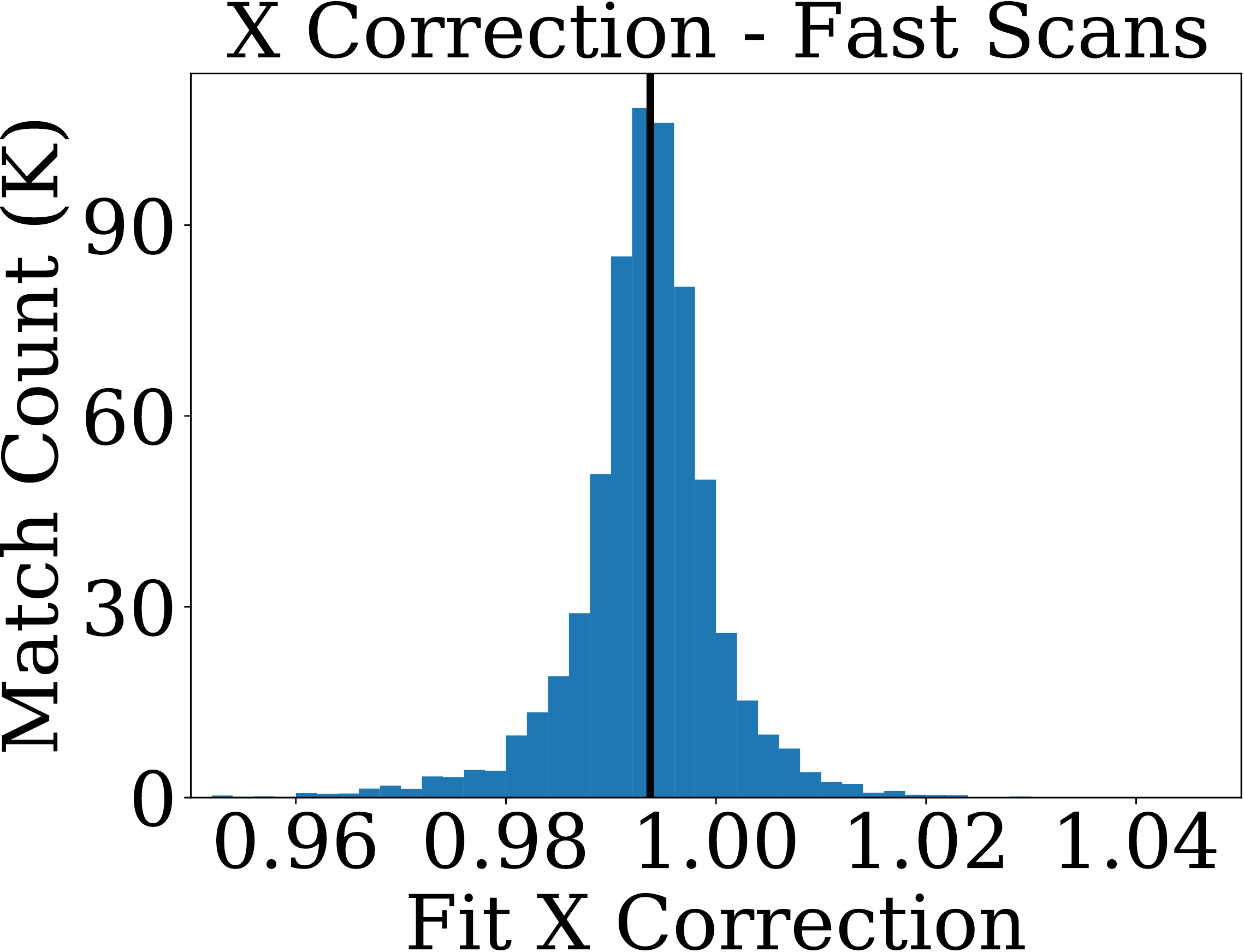} &
    \includegraphics[width=0.48\linewidth]{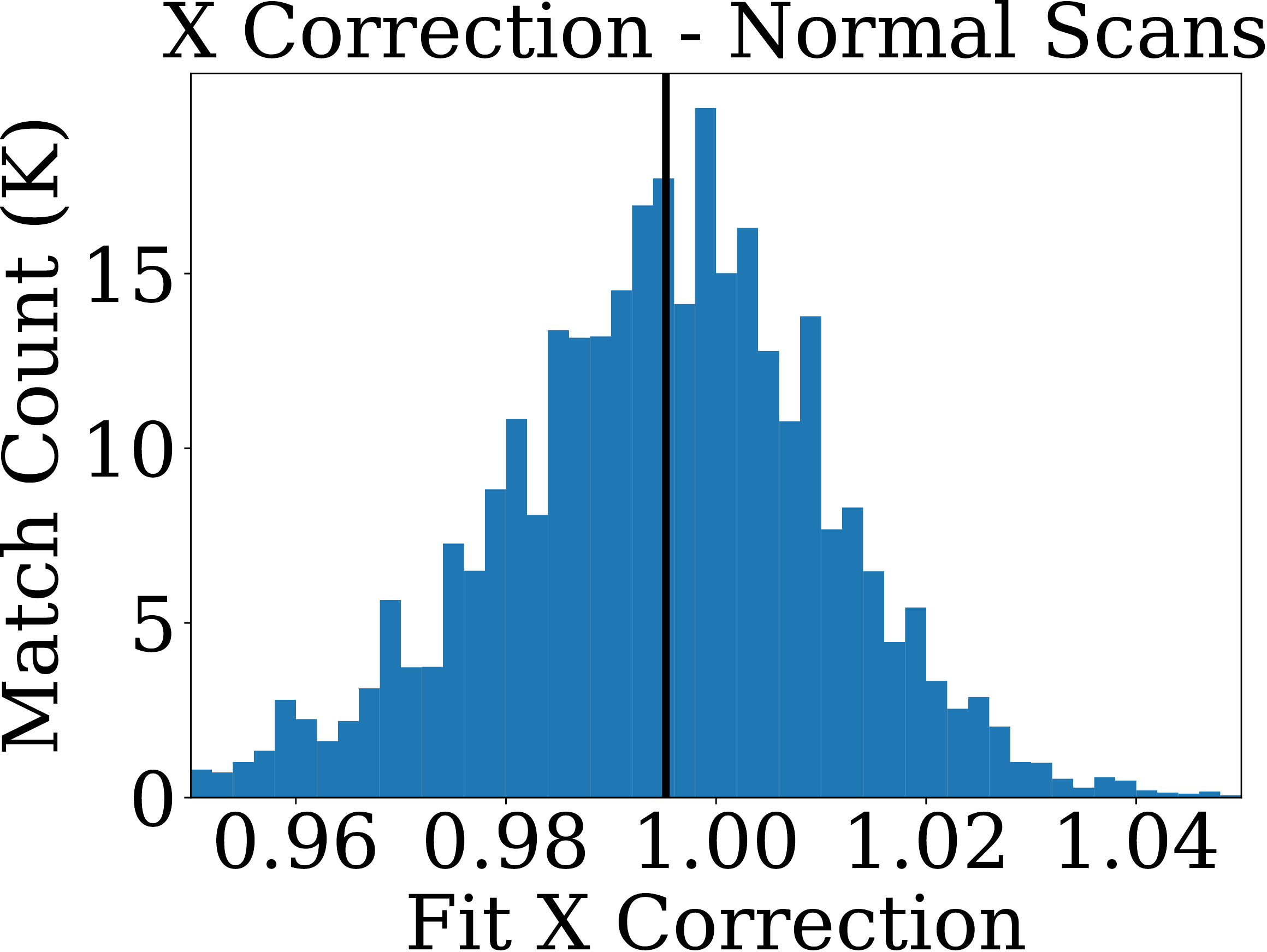} \\
    \includegraphics[width=0.48\linewidth]{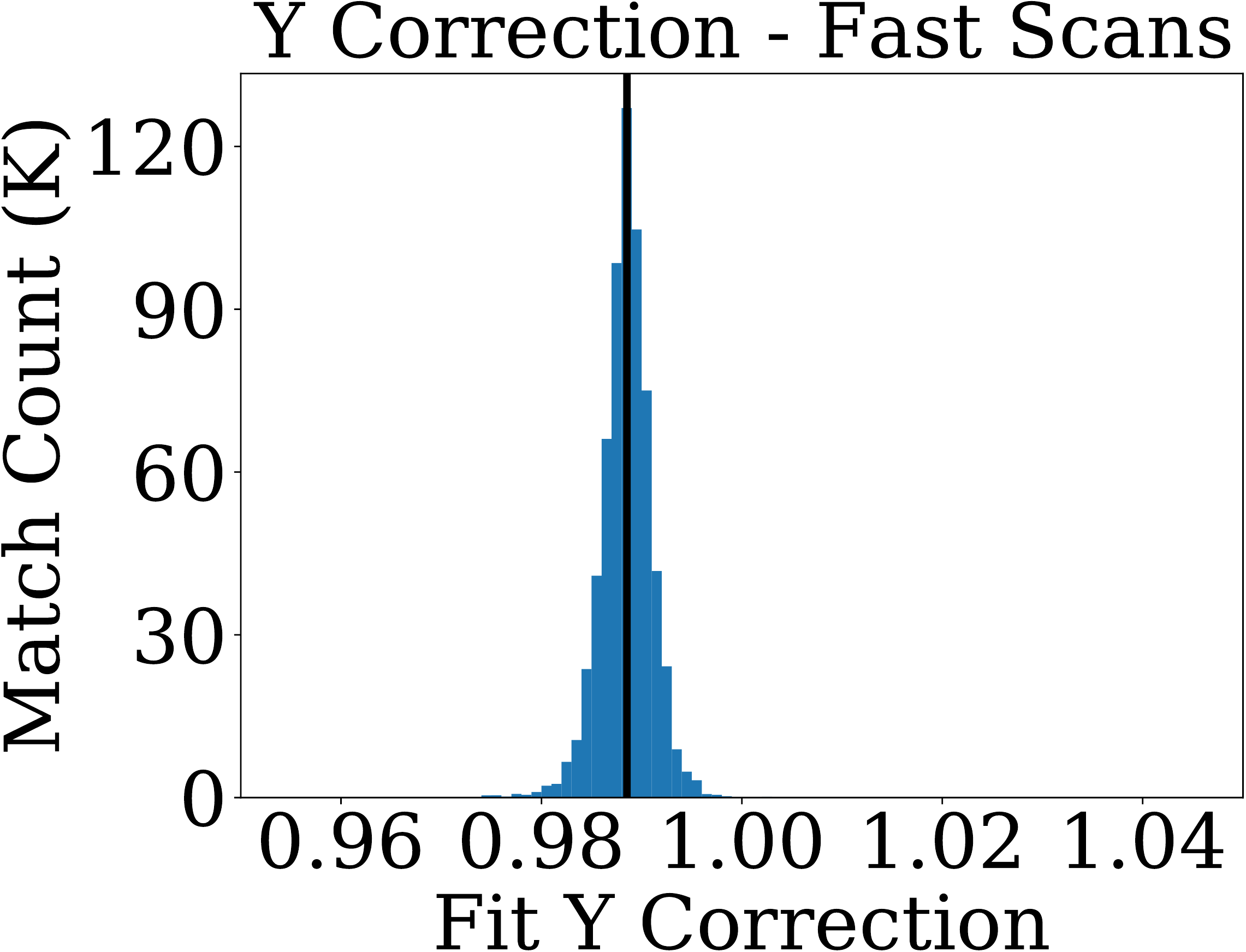} &
    \includegraphics[width=0.48\linewidth]{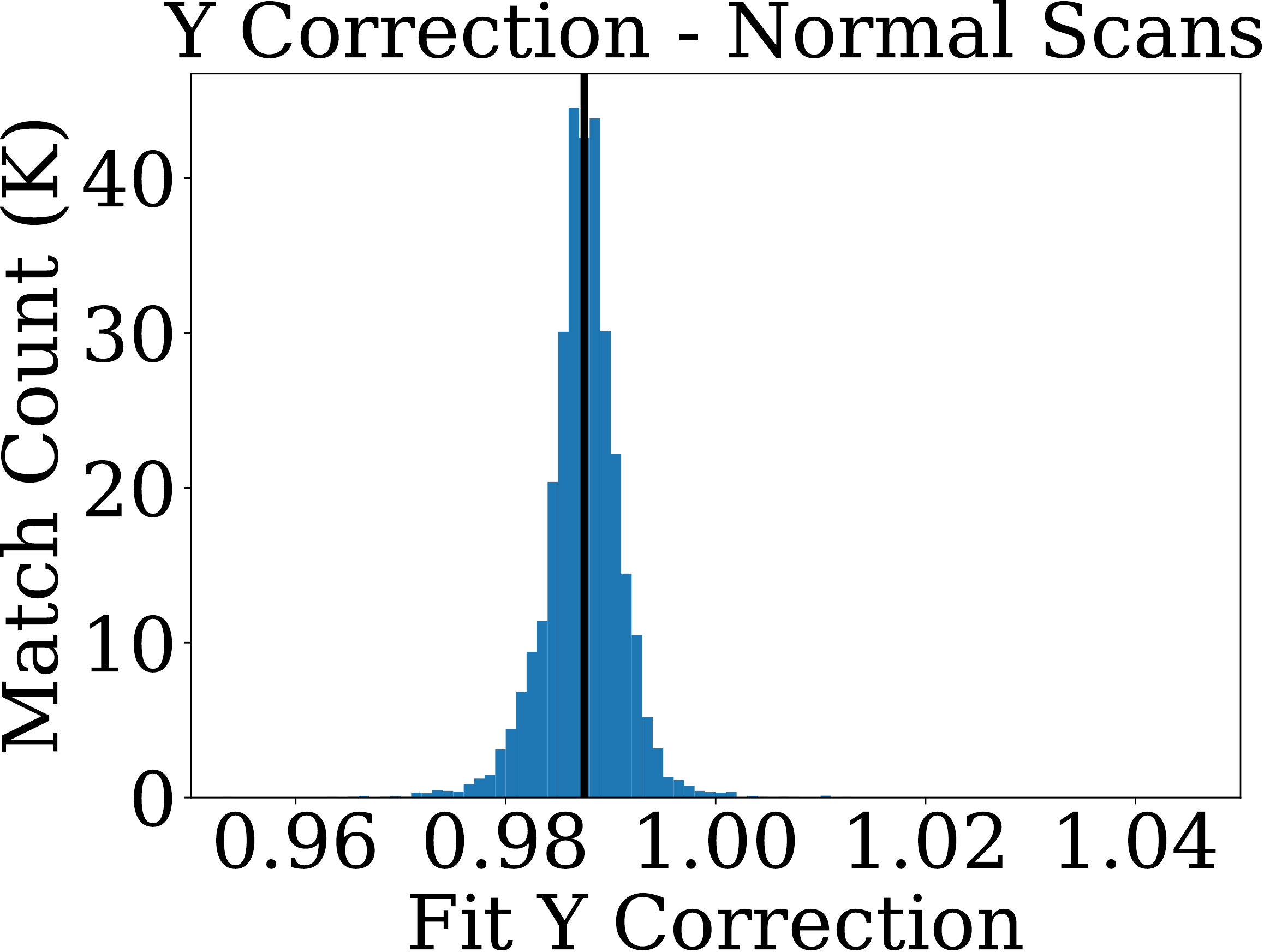} \\
\end{tabular}
    \caption{{\bf Histograms of the fit scale correction factor, weighted by the number of inlier matches.} The 25\% trimmed mean is drawn on the histogram as a black vertical line. There is a strong preference for X and Y scalings that are less than the theoretical Level 2 value on both normal and fast scans. There is substantially more scatter in the X direction, especially for normal scans.}
    \label{fig:scale_histograms}
\end{figure}

We begin by identifying the relative scale of the two instruments' observations. We first treat the scales $s_x$ and $s_y$ as a parameter to fit over the entire dataset, and report an estimate of their values as described by the correspondences. We then analyze whether the updated scales can better explain the data compared to the theoretical scale. Having fit the data, we examine whether there are spatiotemporal trends in the scales that were fit. We additionally consider and exclude some alternate  explanations for our results. Throughout this analysis, we use the Fast and Normal subsets of the data; the full dataset is used for pointing information.

\subsection{Estimating Fitting}

\begin{deluxetable}{lcccccc}[t]
\caption{{\bf Fit scales (in arcseconds) compared with Level 1 and Level 2 reported scales.} The Level 1 y scale is from {\tt CDELT2} and the x scale is the nominal stepping size scaled appropriately (1x for normal scans and 2x for fast scans). The fit scales match the Level 1 data  better than the Level 2 data.}
\label{tab:fitreported}
\tablehead{
\nocolhead{} & \multicolumn{3}{c}{X Scale} & \multicolumn{3}{c}{Y Scale} \\
\colhead{Scan} & \colhead{Fit} & \colhead{L1} & \colhead{L2} & \colhead{Fit} & \colhead{L1} & \colhead{L2}}
\startdata
Fast & 0.2953 & 0.2952 & 0.2971 & 0.3163  & 0.3170  & 0.3199
\\
Normal & 0.1479 & 0.1476 & 0.1486 & 0.1580 & 0.1585 & 0.1599
\enddata
\end{deluxetable}

\begin{deluxetable}{lcccccc}[t]
\caption{{\bf Fit Scales.} In the top two rows, we report percent of the nominal Level 2 value. In the bottom two rows, we report the updated scales in arcseconds.  We report the 25\%-trimmed mean of the scales fit via the full model, the affine model (that does not assume a particular rotation center), and their difference. The affine model results in a similar conclusion, that the scale is smaller than is reported. }
\label{tab:fitscales}
\tablehead{
\nocolhead{~} & \multicolumn{3}{c}{X Scale} & \multicolumn{3}{c}{Y Scale} \\
\nocolhead{Scan} & Full & Aff. & Diff. & Full & Aff. & Diff. 
}
\startdata
Fast & 99.37 & 98.95 & 0.43 & 98.85 & 98.45 & 0.41
\\
Normal & 99.52 & 99.21 & 0.32   & 98.75 & 98.50 & 0.25
 \\
 \midrule
Fast & 0.2953 & 0.2940 & 0.0013  & 0.3163 & 0.3150  & 0.0013
\\
Normal & 0.1479 & 0.1474 & 0.0005  & 0.1580 & 0.1577& 0.0004 
\\
\enddata
\end{deluxetable}

We first fit the full model (which has scaling as a free parameter) on the {\it Fast Subset} and {\it Normal Subset} dataset. Some fits fail, and therefore we report the 25-\% trimmed mean (i.e., the mean of the middle 50\% of the data). We found the results to be insensitive to whether we took the mean scale, weighted by correspondence or not, and to various trim-levels. We plot histograms of the fit scales in Figure~\ref{fig:scale_histograms}, weighted by the number of matches, and plot the 25\%-trimmed mean. In the Y-direction, both fast and normal scans show a substantial preference for a scale at near 98.8\% the theoretical Level 2 value, with little scatter. In the X direction, there is substantially more uncertainty, but a preference for scale near 99.4\% of the theoretical Level 2 value. We report the fit scales compared to the Level 1 and Level 2 reported values in Table~\ref{tab:fitreported}. The data show substantially better agreement with the data from \hinode Level 1: fitting finds a value that is 100.2\% of the nominal Level 1 value in Y and 99.7\% of the nominal Level 1 value in X.

To exclude the possibility that the full model is poorly specified e.g., that something is wrong with the rotation location -- we analyze the inferred scales of the affine model. We factor the fit affine models into a rotation, scaling, and rotation via the SVD, and report the fit scales in Table~\ref{tab:fitscales}. The results are virtually identical, with relative changes in scale of at most 0.41\%. Moreover, the affine model consistently estimates a {\it smaller} scale factor.

To help quantify the adjustment of scale factors, we convert it to pixels. An adjustment of 1.2\% to the Y scale factor of amounts to ${\approx}3.8$px error over a typical 320 \hmi pixels that a \hinode scan covers when warped to the \hmi grid. However, with a model that is fit to describe a mapping from a \hinode scan to a smaller cutout on \hmi, some of the error can be compensated for by adjusting the translation of the center of the window. This leads to ${\approx}1.9$px error over a typical \hmi window. This calculation suggest that \hmi pixel-sized errors can cause a scale adjustment of ${\approx}0.625\%$.

\subsection{How well do the data fit?}

We then turn to testing how well models explain the data when they assume different relative scales. While the data are fit to correspondences, once the transformations models have been fit, they can be used to warp the full \hinode scan to the \hmi grid. The warped \hinode data can then be compared with the corresponding \hmi grid, and its agreement can be used as a measurement of goodness of fit. In particular, we compare a model that assumes the theoretical scales given by the Level 2 header keywords (the {\it Theoretical} model) and one that assumes the empirically fit scale with the corrections found in this paper (the {\it Empirical} model).

\begin{deluxetable}{cccc}[t]
\caption{{\bf Comparison of the Spearman $\rho$ rank correlation between \hmi data and \hinode data, warped with either an Empirical model, or the Theoretical Level 2 model.} Results presented average over field, inclination, and continuum, but trends hold for each individual quantity. Ties are defined as within the 5\% of the the average rank correlation.}
\label{tab:rankcorr}
\tablehead{
\colhead{Scan Type} & 
\colhead{Empirical Better} & \colhead{Tied} & \colhead{Theoretical Better} }
\startdata
Fast & 58.7 & 41.3 & 0.0 \\
Normal & 27.6 & 72.4 & 0.0 \\
\enddata
\end{deluxetable}

\begin{deluxetable}{lcccccc}[t]
\caption{{\bf Comparison of the extent of the features that are well-explained for the empirically fit and theoretical Level 2 parameters.} We report the fraction of times the empirical model covers more of each axis by at least 10 pixel ({\it More}); the fraction that the extents are within 10 pixels ({\it Tie}); and the fraction of times the theoretical model explains a larger fraction of the image ({\it Loss}). The empirical correction often explains correspondences across more of the image, and rarely less. Calculated on matched scans where both have at least 20 inliers.}
\label{tab:width}
\tablehead{& \multicolumn{3}{c}{X} & \multicolumn{3}{c}{Y} 
\\
\colhead{Scan} & 
\colhead{More} & \colhead{Tie} & \colhead{Less} & 
\colhead{More} & \colhead{Tie} & \colhead{Less}}
\startdata
Fast & 
56.5 & 
38.6 & 
4.9 & 
88.3 & 
10.6 & 
1.2 \\
Normal & 
17.3 &
73.8 & 
8.9 &
73.4 & 
24.5 & 
2.0 
\\
\enddata
\end{deluxetable}

\begin{figure*}[t]
\centering
\begin{tabular}{@{}c@{~}c@{~}c@{~}c@{~}c@{~}c@{}}
\multicolumn{3}{c}{Fit X Correction} & 
\multicolumn{3}{c}{Fit Y Correction} \\ 
\includegraphics[height=1.4in]{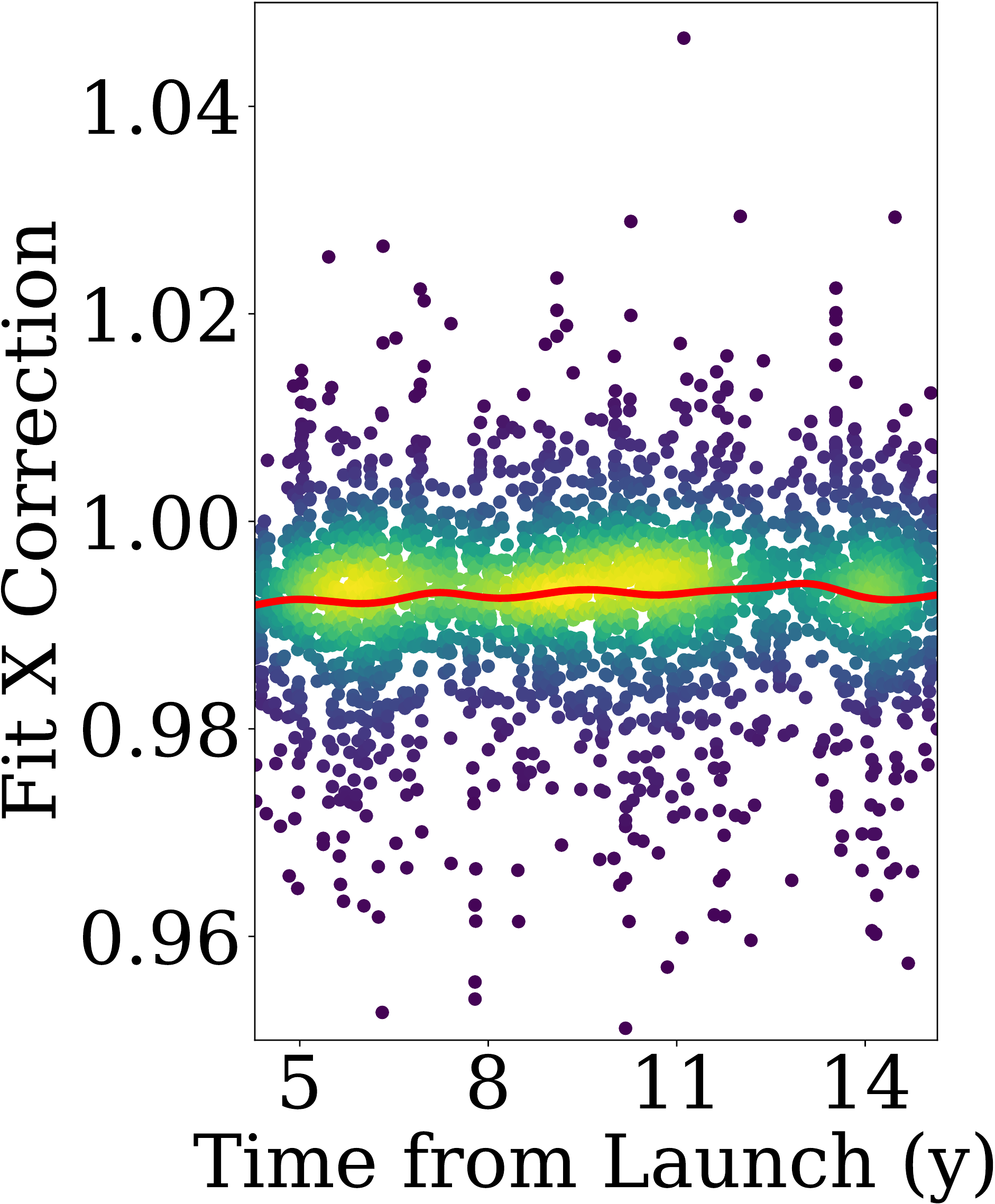} &
\includegraphics[height=1.4in]{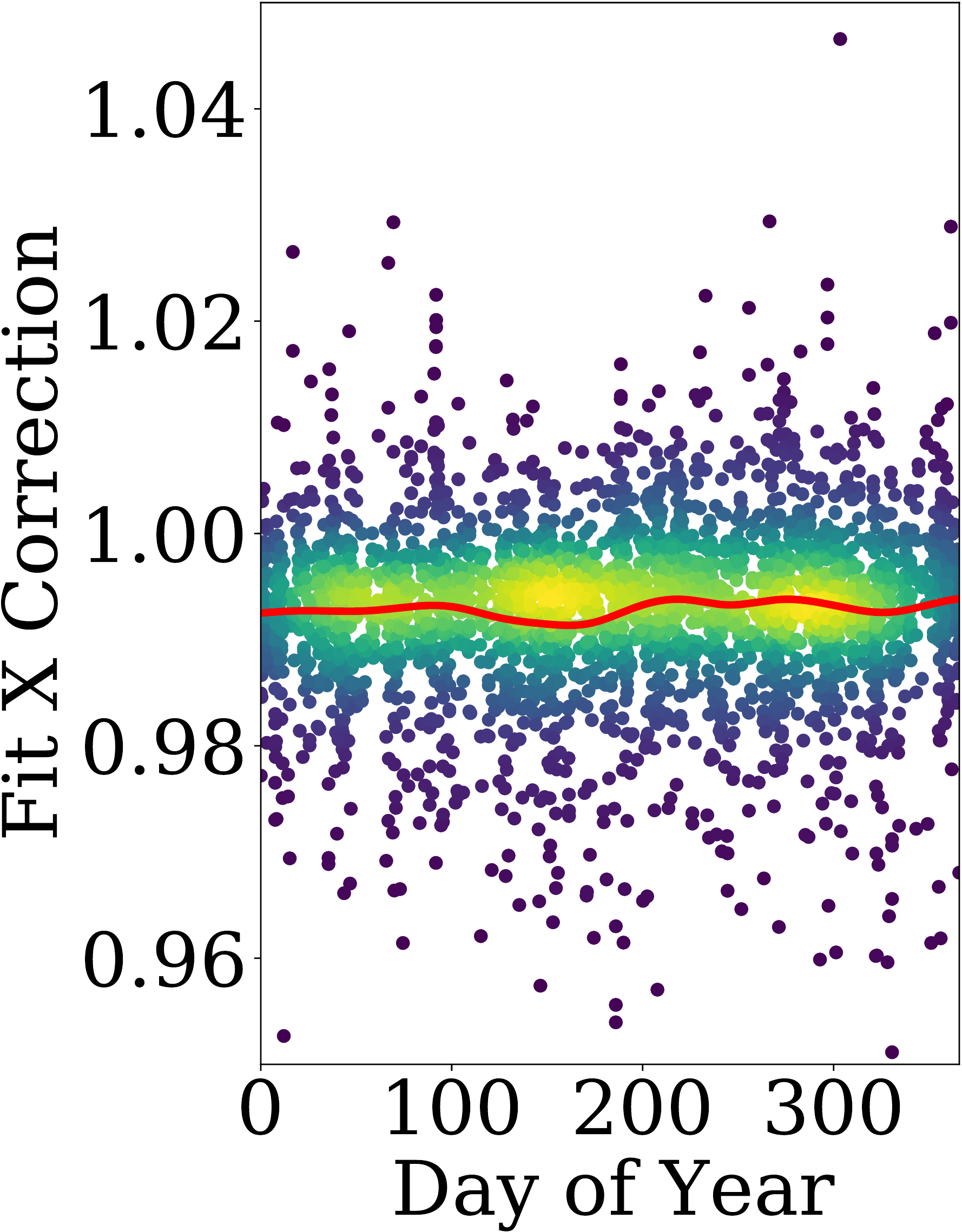} &
\includegraphics[height=1.4in]{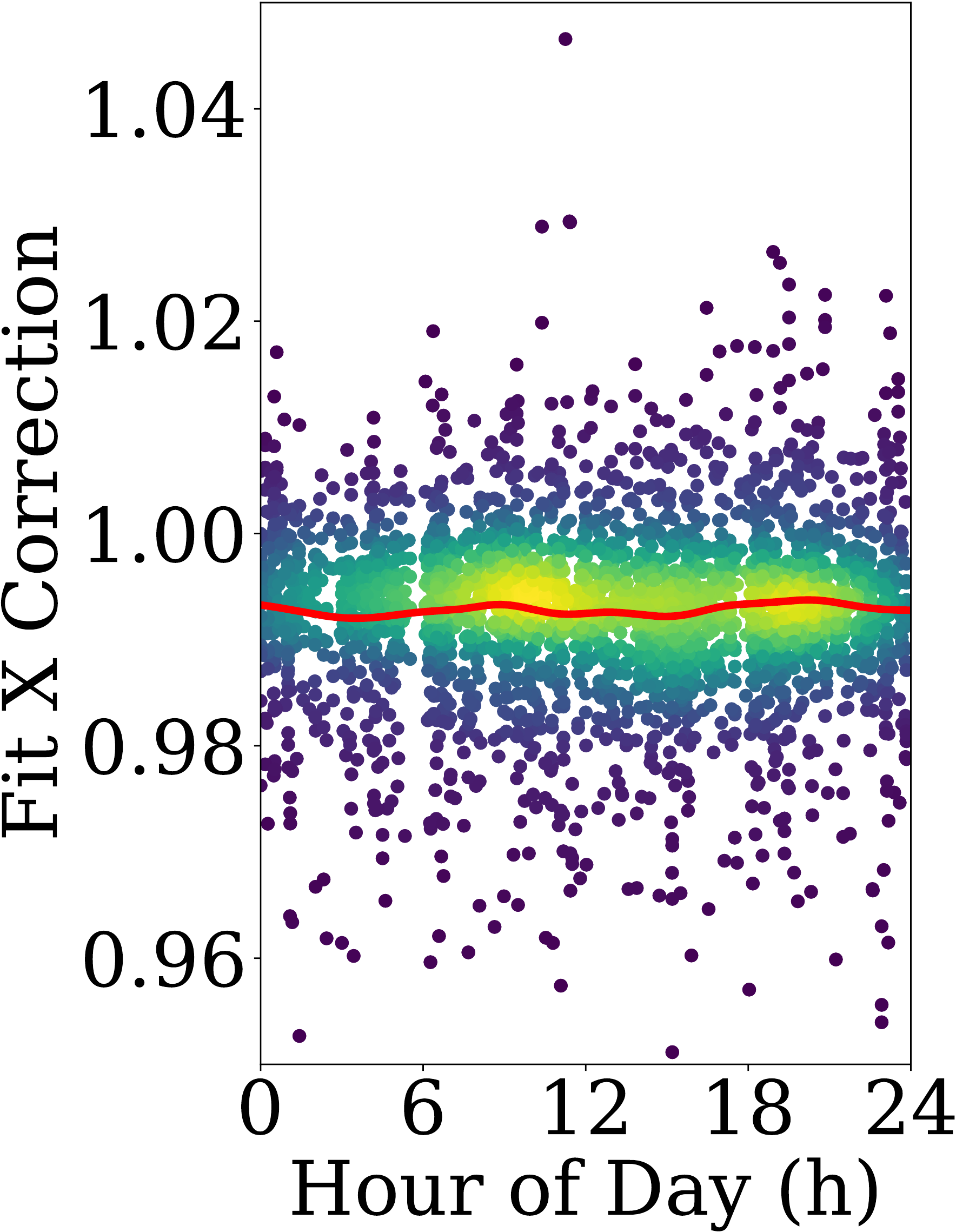} &
\includegraphics[height=1.4in]{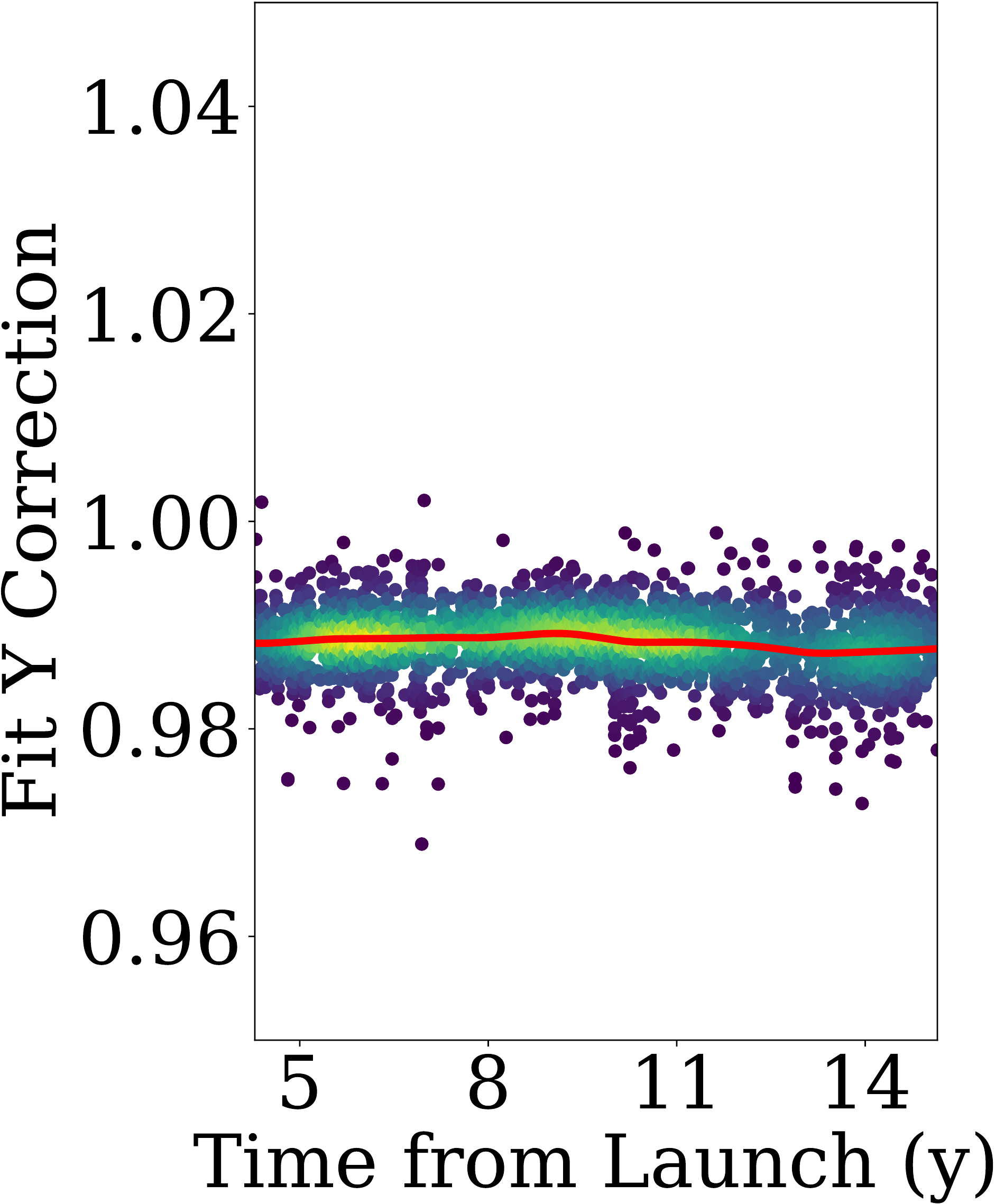} &
\includegraphics[height=1.4in]{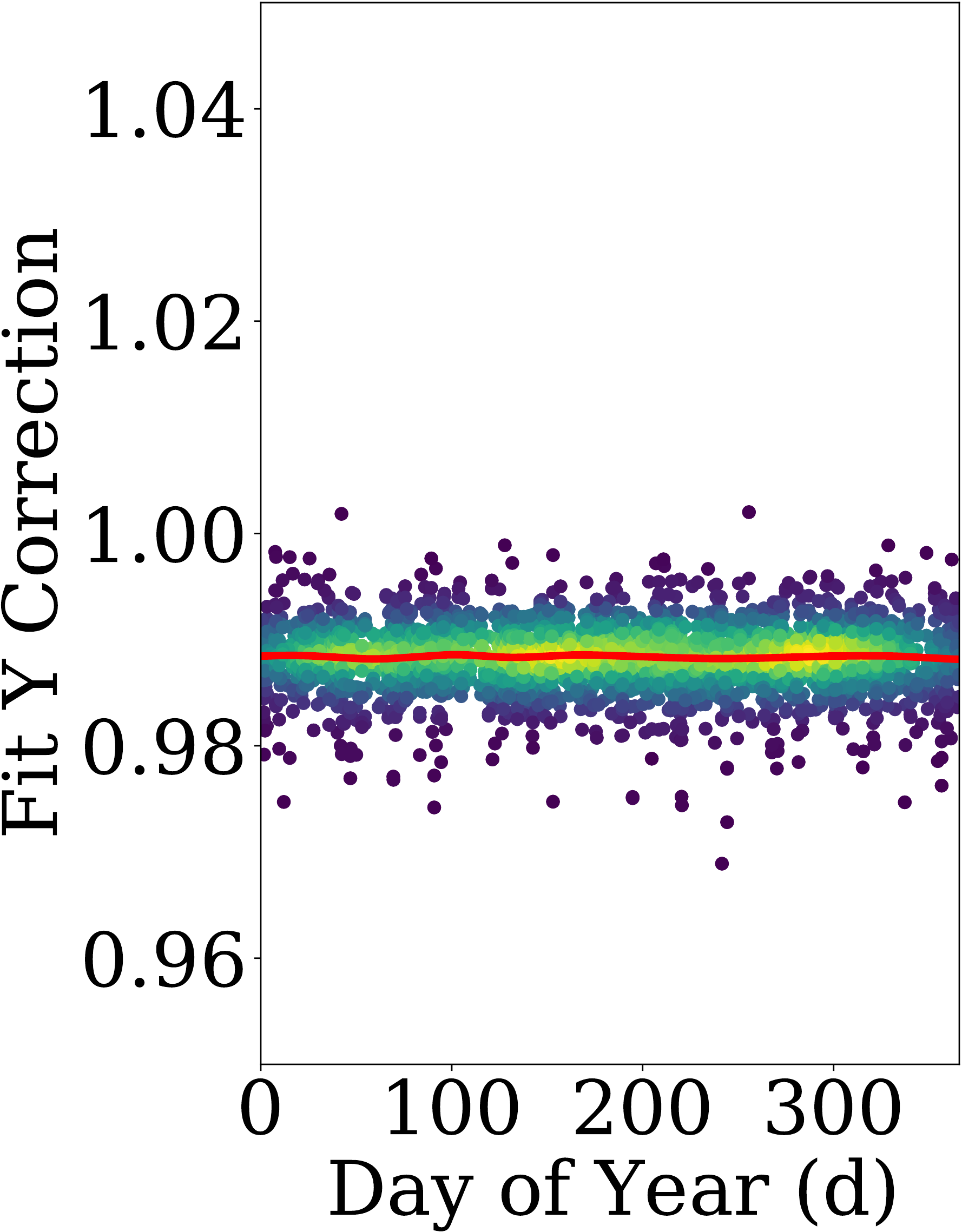} &
\includegraphics[height=1.4in]{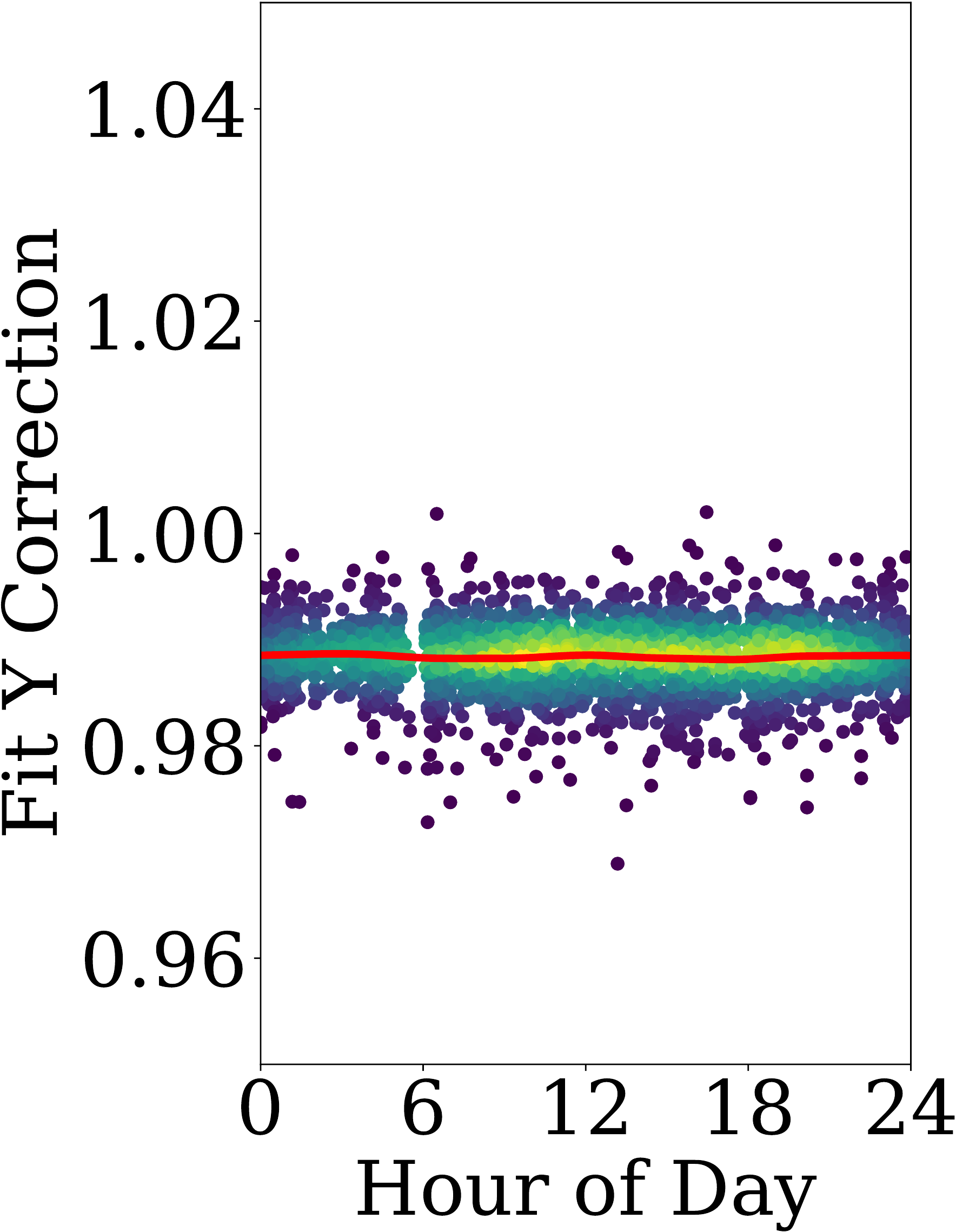} \\
\end{tabular}
\caption{{\bf Scatterplots of fit X and Y corrections against time since launch (22 September 2006), day of year and time of day.} The color of each scatterplot point indicates the density at that location, estimated with a kernel density estimate. We additionally plot a kernel regression estimate in red of scale correction as a function of each covariate. The corrections have no trend, suggesting that the scale factor is constant and that causes driven by Earth or satellite orbit are not responsible.}
\label{fig:temporaltrends}
\end{figure*}

Throughout, we produce the \hinode data warped onto the \hmi grid by interpolating with a degree 3 bivariate spline. As previously described, most \hinode scans have non-uniform slit motion and therefore have missing scanlines. We interpolate these values by linearly interpolating between the recovered scanlines; while a poor approximation over large discontinuities, the data deliberately only includes slits with minor discontinuities.

We evaluate the rank correlation between \hmi data and the warped \hinode data, comparing continuum, inclination, and field strength (\hmi: {\tt field}; \hinode: the product of {\tt Field\_Strength} and {\tt Stray\_Light\_Fill\_Factor}). In this case, since we are doing pixel-to-pixel comparisons, we explicitly compute magnetic flux density for \hinode. Using rank correlation as a measure of agreement compared to e.g., distance, avoids issues of relative calibration and scaling of the two pipeline quantities. While the rank correlation itself is not informative as an absolute quantity, the relative rank correlations can be used to sort different methods: one can check how many times one transformation yields data with a higher rank correlation compared to another.

We use this approach to compare the fits produced with the theoretical Level 2 model and the empirical model. We average over the three quantities and for each \hinode scan, we pick at random from the HMI transformations for which the model has at least 20 inliers. Table~\ref{tab:rankcorr} shows that the empirical correction results in warped data that is never worse than the theoretical values, and often better. 

Another method of testing goodness of the fit scale is quantifying the extent to which inliers for each model cover the image. We quantify performance by examining the spatial extent of the inliers of the theoretical as well as the empirical model. A mis-specified scaling can be corrected for in a small region by adjusting the translation. However, changes in translation cannot compensate over a large region. 

We quantify the extent via the size of the middle 95\% of the data for a set of inliers $[x_i, y_i] \leftrightarrow [x'_i, y'_i]$. For $x$, this is  $\textrm{quantile}(\{x_i\},97.5\%) - \textrm{quantile}(\{x_i\},2.5\%)$ and similarly for $y$. We compare the rate at which the empirical model's inliers cover a wider fraction of the image compared to the theoretical model's inliers and report results in Table~\ref{tab:width}. The empirical model usually covers more of the image, suggesting that the empirical model is a better description for more of the image compared to the theoretical Level 2 values. The difference is consistently visible in the Y direction, and larger for fast scans compared to normal scans. We hypothesize that this difference is caused by a difference in the field of view covered by the scans that could be registered: we found that the fast scans that could be registered typically had a larger field of view compared to the normal scans that could be registered. Since the error due to an incorrect scaling increases linearly as a function of field of view, the fast scans likely have more room for improvement with better alignment.

\subsection{Identifying Correlated Signals}

\begin{figure}
    \centering
    \includegraphics[width=\linewidth]{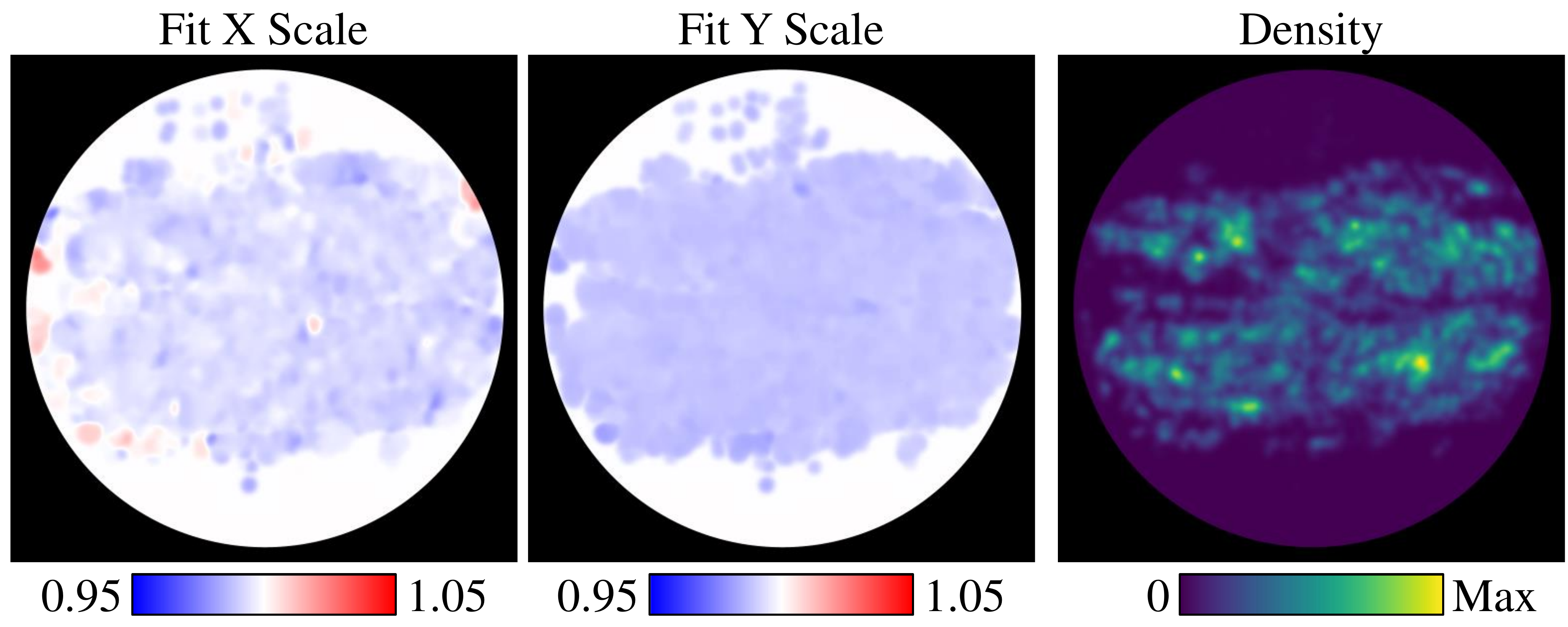}
    \caption{{\bf Fit scales by location.} We plot a kernel regression estimate of the location, with a prior of $1$. We also plot the density. The apparent bias in density towards one side of the disk is driven by the \hinode observing cycle: an active region needs to have become previously visible in order for \hinode to have it selected as a target. The corrections also have no trend (although there are some outliers), suggesting that the location-dependent causes such as differences in line formation height are not responsible for the change in scale.}
    \label{fig:density}
\end{figure}

\begin{figure}
\centering
\begin{tabular}{@{~~~}c@{~~~~}c@{~~~}}
\includegraphics[height=2.0in]{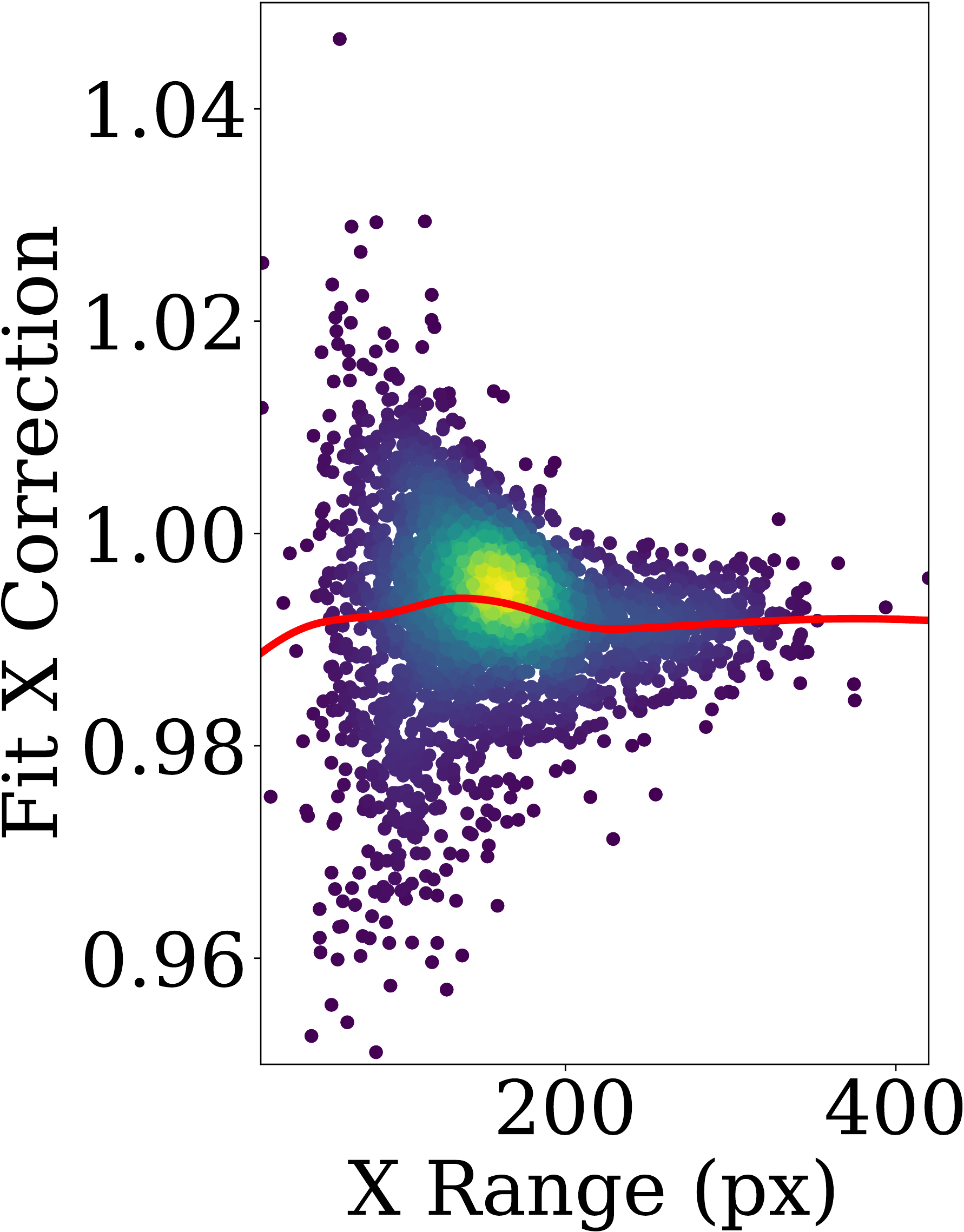} & 
\includegraphics[height=2.0in]{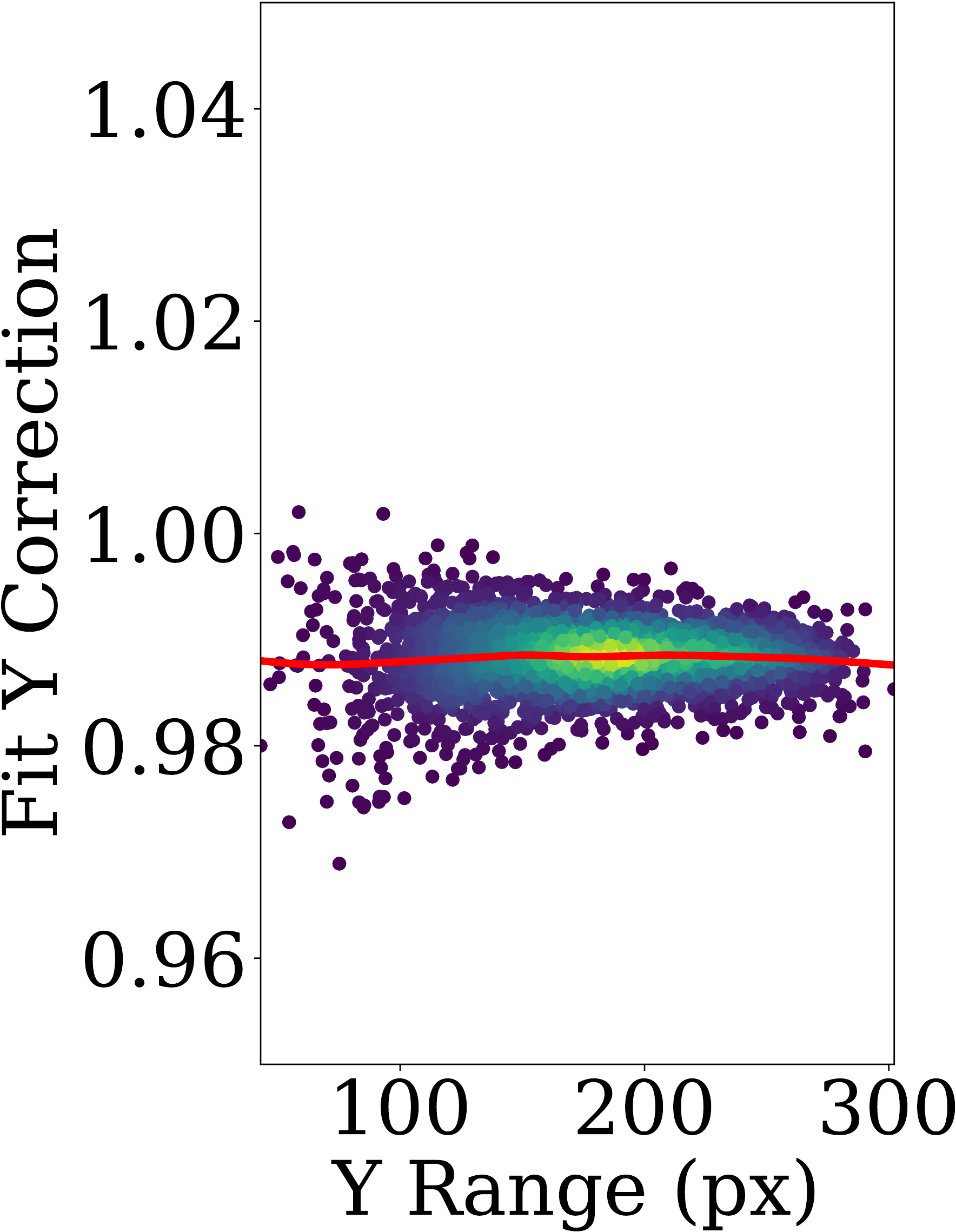} \\
\end{tabular}
\caption{{\bf Scatterplots of fit X and Y corrections against the range of the X and Y coordinates.} Having more correspondences across the full image results in more stable estimate of the scales in the X coordinate.}
\label{fig:stable}
\end{figure}

\begin{figure*}
\centering
\begin{tabular}{@{}c@{}c@{}c@{}c@{}c@{}}
\includegraphics[width=0.2\linewidth]{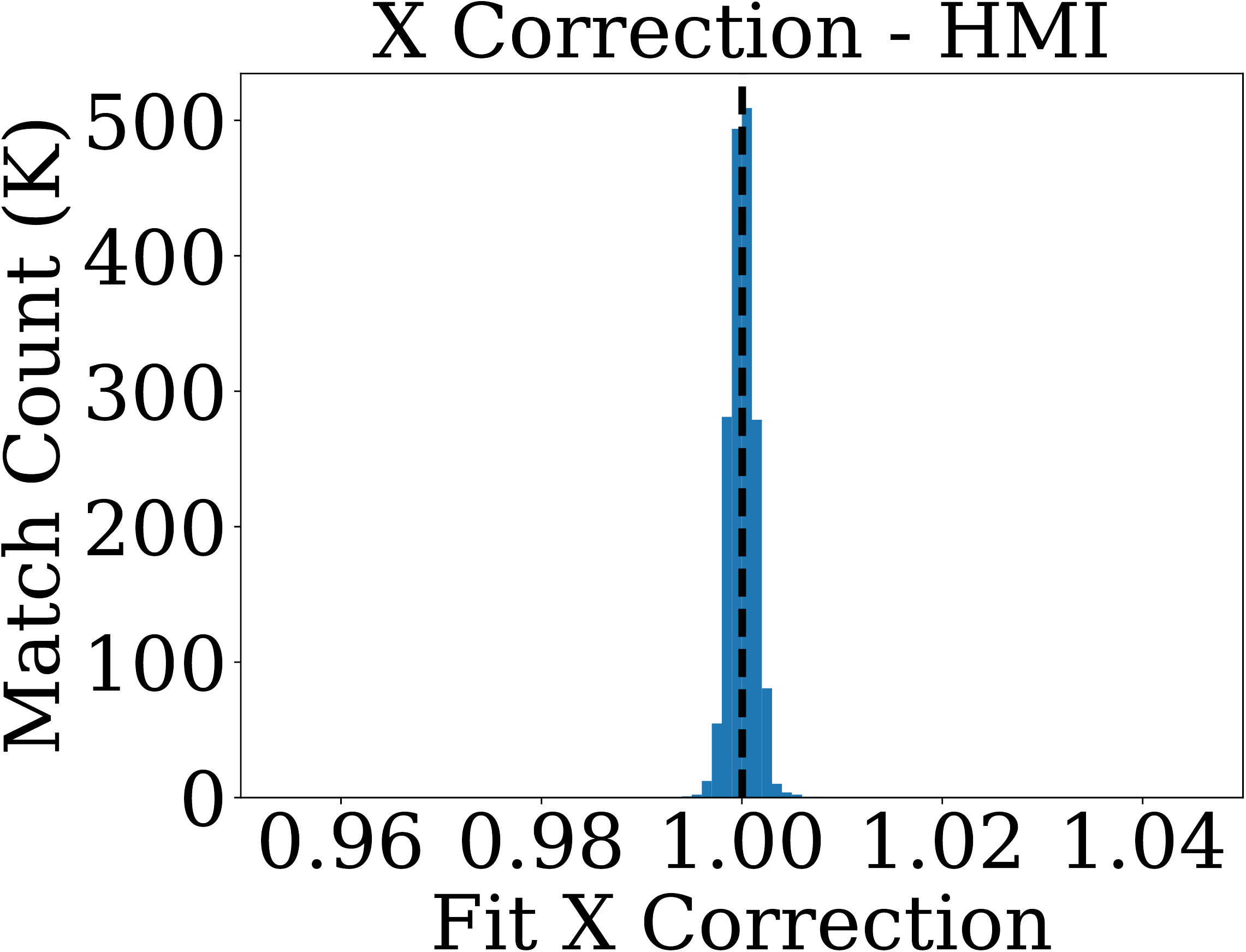} & 
\includegraphics[width=0.2\linewidth]{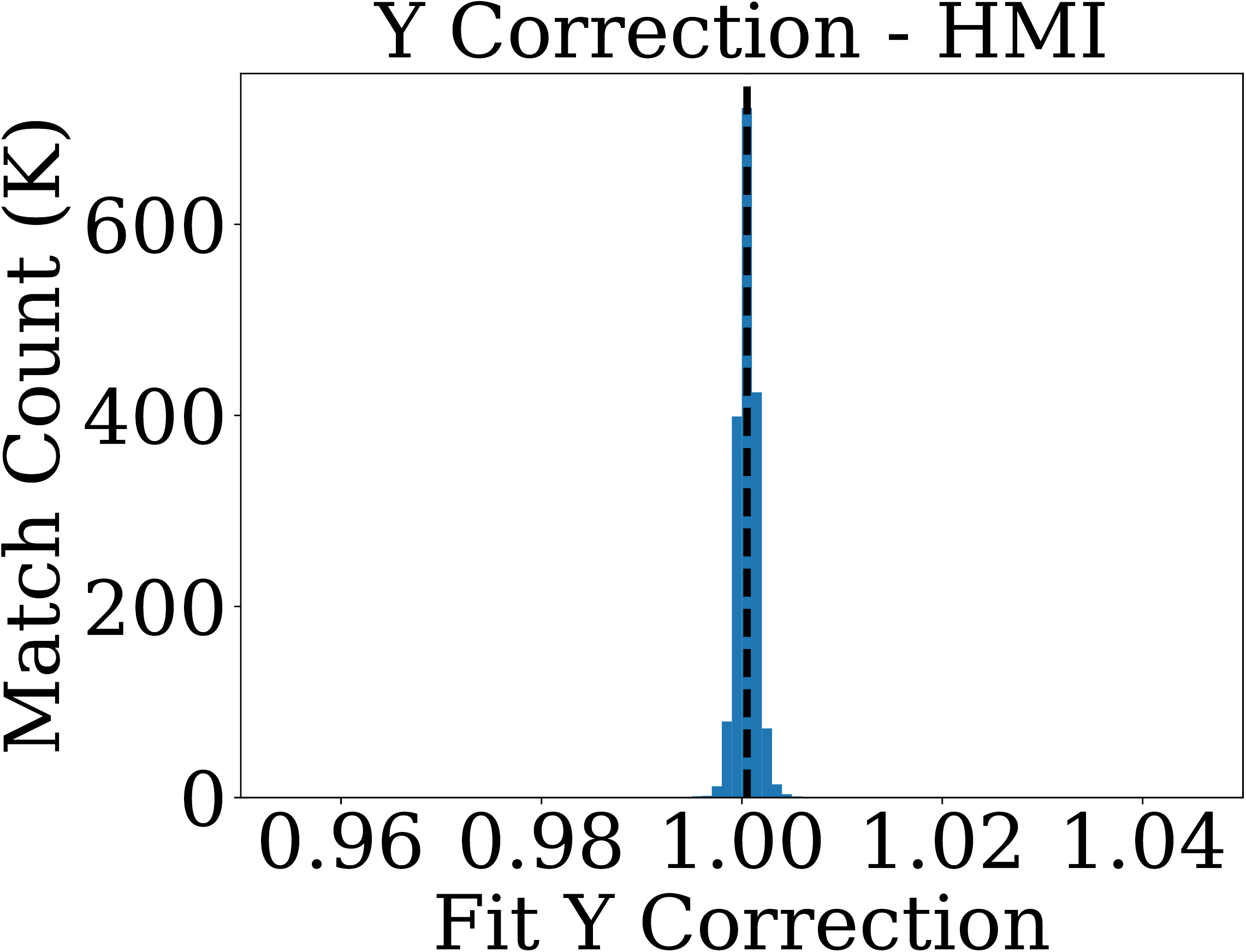} & 
\includegraphics[width=0.2\linewidth]{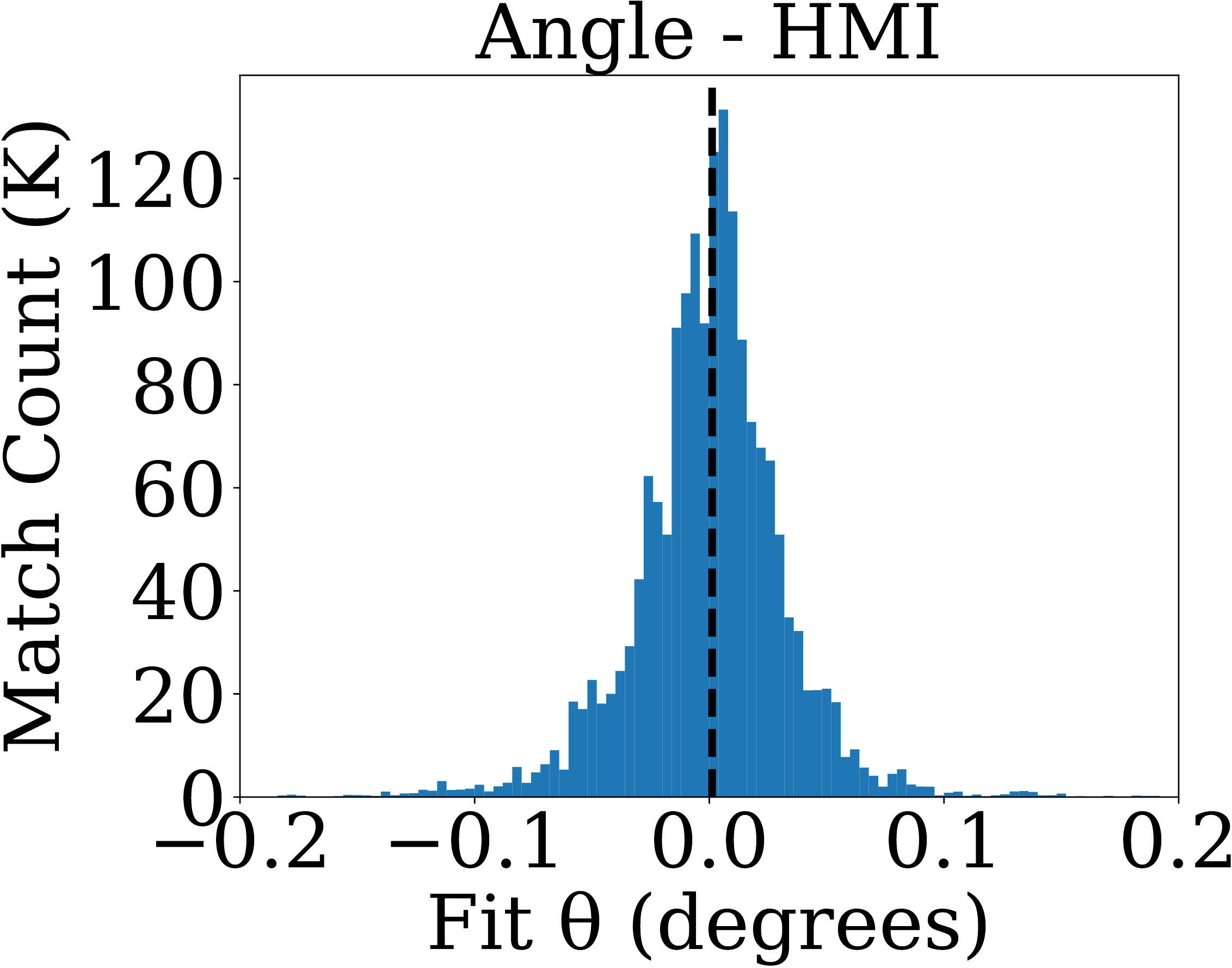} & 
\includegraphics[width=0.2\linewidth]{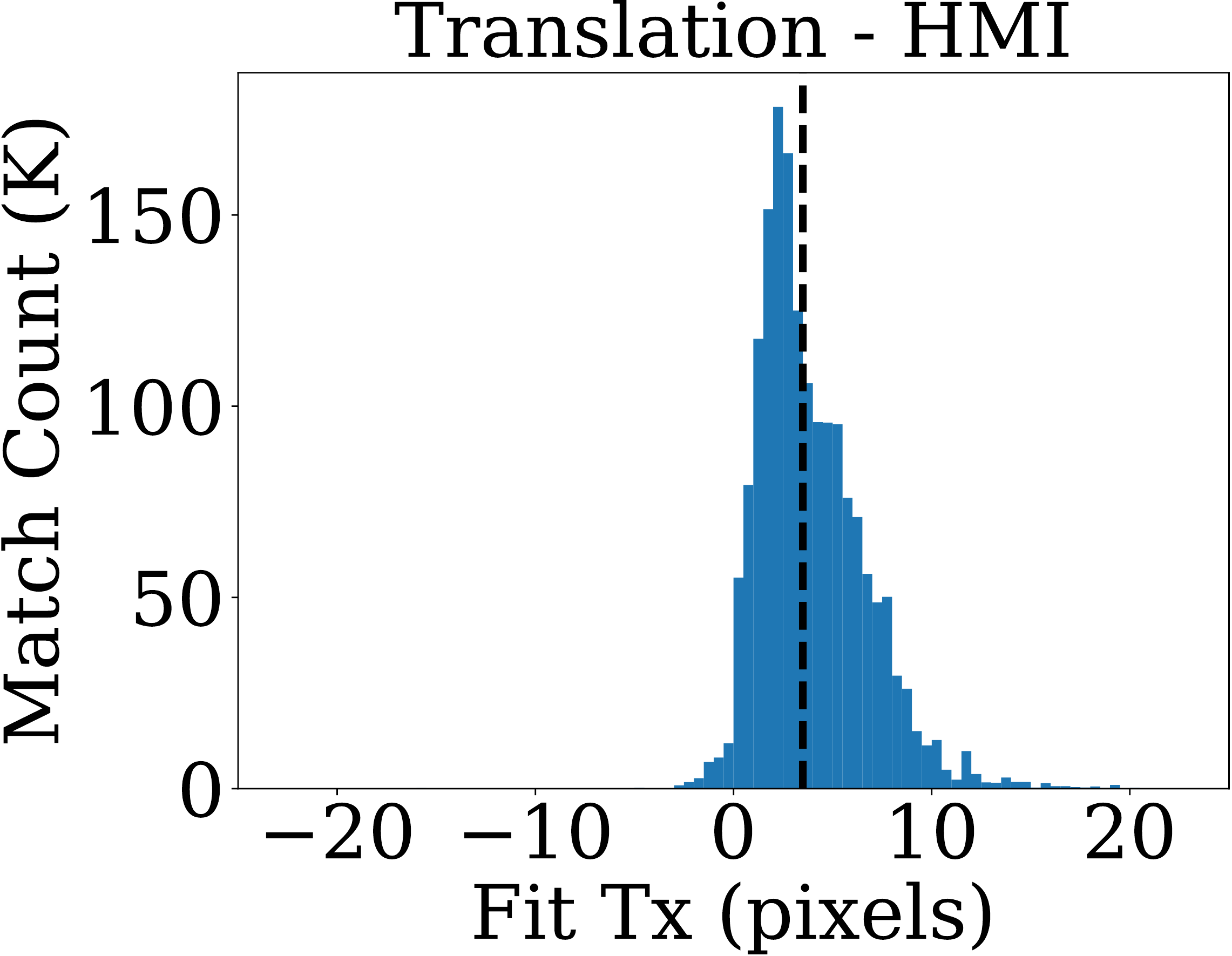} & 
\includegraphics[width=0.2\linewidth]{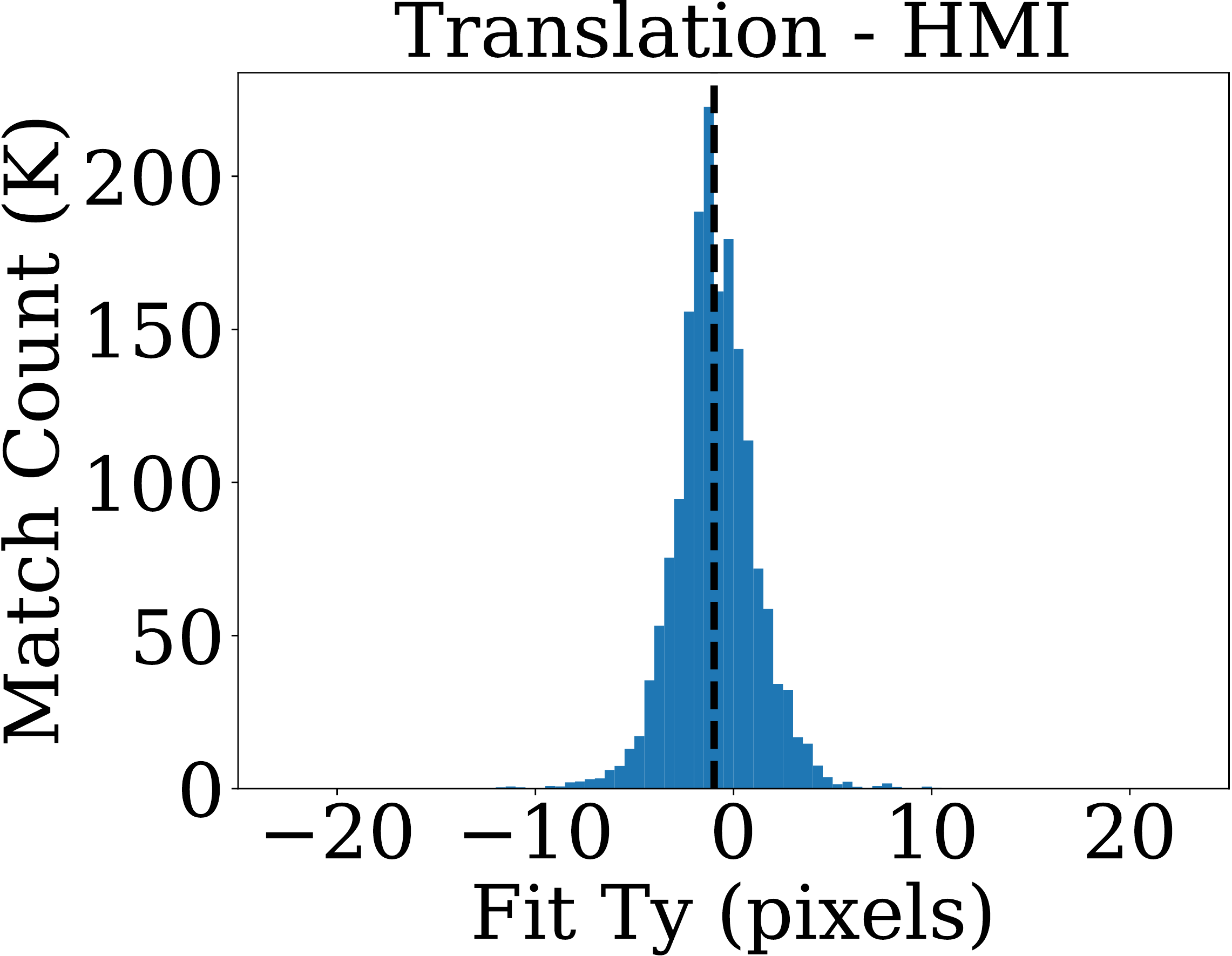} \\
\end{tabular}
\caption{{\bf Fit on temporally adjacent \hmi scans from the {\tt ME\_720\_fd10} series.} In both cases, the fit scales are effectively unity and rotation zero. Translations have substantial scatter with the x axis centered on a positive value, and the y axis centered on zero. This agrees with intuition, suggesting that the fitting procedure is effective even with temporal evolution.}
\label{fig:hmifit}
\end{figure*}

We next turn to identifying whether the estimated scale correlates with temporal or spatial quantities.

For temporal trends, we examine both a secular trend in scale over a decade (similar to the instrument degradation in e..g, \aia's data) as well as cyclic trends in the data. In particular, we test yearly cycles (to account for artifacts related to the Earth's orbit) and daily cycles (to account for artifacts related to orbit such as those in \hmi's  data~\citep{Hoeksema2014}. We plot the fit scale corrections against these variables in Figure~\ref{fig:temporaltrends} along with a kernel regression estimate of the fit scale as a function of the covariate. Given a set of $N$ locations $\{\lB_i\}_{i=1}^N$ 
with values $\{v_i\}_{i=1}^N$, and a kernel function $k$, the value of the kernel regression at a query point $\qB$ is the values weighted by the kernel distance to the query point, or
$(\sum_{i=1}^N k(\qB, \lB_i) s_i) / (\sum_{i=1}^N k(\qB, \lB_i))$. This kernel regression smooths out the noise in each individual fitting. We find no trends in any of the temporal data. This lack of trends would suggest that the pixel scale has not changed over time, which agrees with the intuition that pixel scale is an unchangeable property (at least at the precision we can measure).

Next, we examine the estimated scale as a function of image plane location in Figure~\ref{fig:temporaltrends} to see if there are spatial trends across the disk (e.g., a hemispheric bias or correlation with distance-from-center). We compute a kernel regression estimate, using the HMI coordinates as locations and the fit scales as values. We use a Gaussian kernel with $\sigma=64$px, or $k(\xB,\yB) = \exp(-||\xB-\yB||_2^2 / \sigma^2)$. Many parts of the disk are far from any data point, so for regularization, we also add a single dummy correspondence at the query location with scale 1, leading to a scale of $1$ where there is no data. The data are relatively consistent across the disk. While some spots on the disk deviate from the overall average, these are likely due to a handful of scans, rather than systematic biases.

Finally, we plot the fit corrections against the range of the inlier correspondences in $x$ and $y$ in Figure~\ref{fig:stable}. We define the range via the range of the middle 95\% of the data. Unsurprisingly, as feature correspondences cover less of the image, the correction that is fit exhibits substantially more variability.

\subsection{Analysis of the Correspondences}

For fast data, the bulk of the inlier matches come from 
Inclination (39.1\%) and Field (36.6\%), with continuum (17.2\%), and V channels (7.0\%) making up most of the rest. Most (83.7\%) of the correspondences come from SIFT, with the rest from ORB (although we note that the SIFT detector configuration we used produced more correspondences to start with). Among the image pairs where there are at least 20 inlier correspondences, on average 6\% of the data are inliers and 94\% outliers. This fraction is driven by both the challenging nature of the corresponce problem as well as the fact that a relatively small cutout from \hinode is being matched to a much larger full disk image from \hmi.

The high outlier fraction (94\%) seems daunting at first. However, two things facilitate the extraction of good models. First, the outliers are random, as opposed to structured. This means that RANSAC~\citep{Fischler81} cuts through them with relative ease. Second, although our models that fit rotation and scale require 3 correspondences to be fit, we use a fast pre-screening model that fits to a {\it single} correspondence while using an assumed rotation and scale. This is only possible because $\theta = 0$ is a good approximation for reality and the reported scale parameters are already quite accurate. This insight has also been applied in the context of field robotics by~\cite{civera20101}, where vehicle motion is used to constrain the search for more complex models.

\subsection{Excluded Explanations}

We now briefly discuss factors that we exclude as potential causes of the observed scale factor. These alternate explanations include the scale factor being used to explain away temporal evolution, the relative locations of both the instruments and the phenomena they observe, and issues with the accounting for the mechanical slit position. We cannot rule out that these factors {\it cause} changes in scale factor because each obviously results in some relative scaling. Instead, we want to demonstrate that each is too small to plausibly explain the observed changes and would have left a signature in our results.

\begin{deluxetable}{lcccccc}[t]
\caption{{\bf Fit Scales for both all our data and near-smooth scan data.} We report scales (in \%) on both all the scans from our fast dataset that can be registered with at least 20 correspondences explained within 1px, as well as the subset of these where the number of excess slit positions is no more than two positions. We report the 25\%-trimmed mean of the scales fit via the full model. We reach the same conclusions when using the smaller number of scans with close-to-no jumps.}
\label{tab:fitscalesStrict}
\tablehead{ & \multicolumn{3}{c}{All Scans} & \multicolumn{3}{c}{$\le2$ Extra Slit Positions} \\
\colhead{Scan} & \colhead{X} & \colhead{Y} & \colhead{Count} & \colhead{X} & \colhead{Y} & \colhead{Count}} 
\startdata
    Fast 
    & 99.37 & 98.85 & 760
    & 99.38 & 98.82 & 287
    \\
    Normal  
    & 99.52 & 98.75 & 503
    & 99.95 & 98.70 & 98
\\
\enddata
\end{deluxetable}

One hypothesis for the observed fits is that solar evolution during the \hinode scan is responsible for the change in scale. If one rotates the coordinate system by $\theta \gg 0$, a change in the y scale can compensate for changes in the x direction. We therefore test the fitting procedure by registering temporally adjacent HMI scans, and observing the fit parameters and rotations. 

The relationship between the pixels in temporally HMI scans {\it across the full image} is not described by any of the models used in this paper. However, locally within an active region, these transformations are a reasonable model. We therefore perform fits on the same active regions imaged by \hinode, which we identify by mapping the \hinode scan warped to the \hmi grid and computing its bounding rectangle. We find correspondences between \hmi's field, azimuth, inclination, and continuum, filtering the correspondences to lie in this rectangle. There is no scaling or rotation between temporally adjacent \hmi scans, and so we should expect $s_x$ and $s_y$ to be near unity and $\theta$ to be near zero. The Sun's rotation ought to produce usually positive translation in the $x$ component. Because the Sun is rotating out of the image plane, the $y$ rotation will not be zero in each image; however, unless there are biases in data sampling, it should have mean zero over the dataset. We report results in Figure~\ref{fig:hmifit}. The fit scale corrections lie close to unity. The Y scale has a 25\% trimmed mean of 100.05\%, and 95\% of the data falls between 99.86\% and 100.24\%. The X scale has a 25\% trimmed mean of 100.00\% with 95\% of the data between 99.77\% and 100.25\%. The fit rotation tends to have a magnitude near zero. Translation has substantial scatter, but $Y$ is centered on zero and $X$ is positive.

Another possible explanation is the relative positioning of the spacecraft carrying the instruments. However, they are insufficiently far apart to explain the change in scale we observe: SDO's orbit is less than ${\approx}3.6\times10^4$ km and Hinode's orbit is less than $7 \times 10^2$ km. The sum of their orbits gives an upper bound on their potential distance, yielding ${\approx}3.7\times10^4$ m, or just 0.025\% of the average Earth-Sun distance ($1.5 \times 10^8$ km). This fraction of the distance is negligible compared to the scale diferences seen (${\approx}1.18$\% and ${\approx}0.64$\%). Additionally, if spacecraft position were a factor, we would expect to see variation in scale as a function of time of day  via the interaction between both spacecrafts' orbits or time of year. The plots in Figure~\ref{fig:temporaltrends} show no such trends.

\begin{figure*}[t!]
    \centering
    \includegraphics[width=\linewidth]{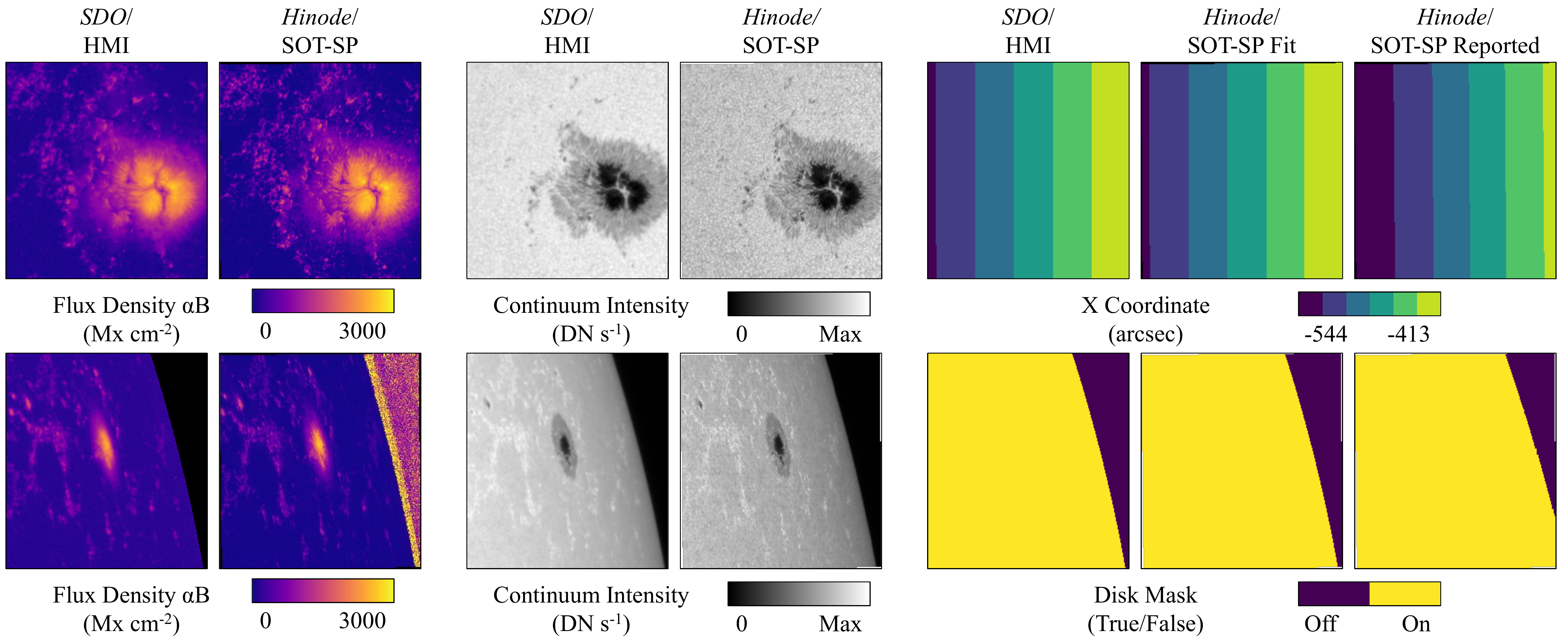}
    \caption{
    {\bf Warped data and pointing coordinates.} We show a subset of the maps for conciseness, including flux density ($\alpha B$) and continuum intensity and coordinate maps and disk masks. Coordinates shown for {\it fit} values that are produced by the fit alignment with \hmi and {\it reported} values that are reported in the Level 2 \hinode data. The reported values show substantial disagreement with \hmi data even though the alignment of continuum and field strength are close to pixel-perfect. This is most clear by examining the limb pixels on bottom. In comparison, our fit corrections align well. Top: 2011 April 30, 09:55:06 TAI. Bottom: 2013 April 3, 15:47:05 TAI.}
    \label{fig:pointing}
\end{figure*}

In addition to being at different locations, the instruments observe different parts of the photosphere. These height differences are also insufficiently different to explain the scale differences.
\cite{Higgins2022} report a theoretical estimate of no more than ${\approx}1.5 \times 10^2$~km difference of formation heights using results given in~\citep{grec2010measuring,norton2006spectral}. This height difference results in at most a sub-pixel misalignment. This upper bound can be computed by making
assumptions that maximize misalignment. We compute the $150$km height difference at the limb and assume the Sun takes up the full sensor, yielding a radius of 2048 HMI pixels. In this setup, each pixel amounts to ${\approx}339$km per HMI pixel. The error due to line height is thus $\le 0.443$px, which is a loose upper bound given that line formation height difference has no impact at disk center. This error, however is substantially smaller than the translation error induced by a scale change of $1.9$px. Moreover, the difference in pixel location due to formation height varies as a function of distance to disk center. If formation height were the cause of scale differences, it would leave a signature of substantial variation of the fit scale factor across the disk. However, the plots in Figure~\ref{fig:density} show a near constant function for $Y$ and while $X$ is noisier, no consistent trends emerge.

We finally consider jumps in the scanning slit mechanism. While we account for these in our fitting, we re-run the analysis on a subset of scans where there are nearly no jumps. In particular, we consider only scans with two or fewer extra slit positions, corresponding to $0.297''$. For \hmi, this amounts to ${\approx}0.5$ pixels. For \hinode, it is one pixel in fast mode, and two pixels for  normal mode. We report results in Table~\ref{tab:fitscalesStrict}. Our results show good agreement with the full dataset with the exception of the scale fit for Normal scans in the X direction. This quantity is, however, both noisy and fit on a small number of scans.

\section{Estimating Relative Pointing}
\label{sec:pointing}

\begin{figure*}[t]
\centering
\includegraphics[width=\linewidth]{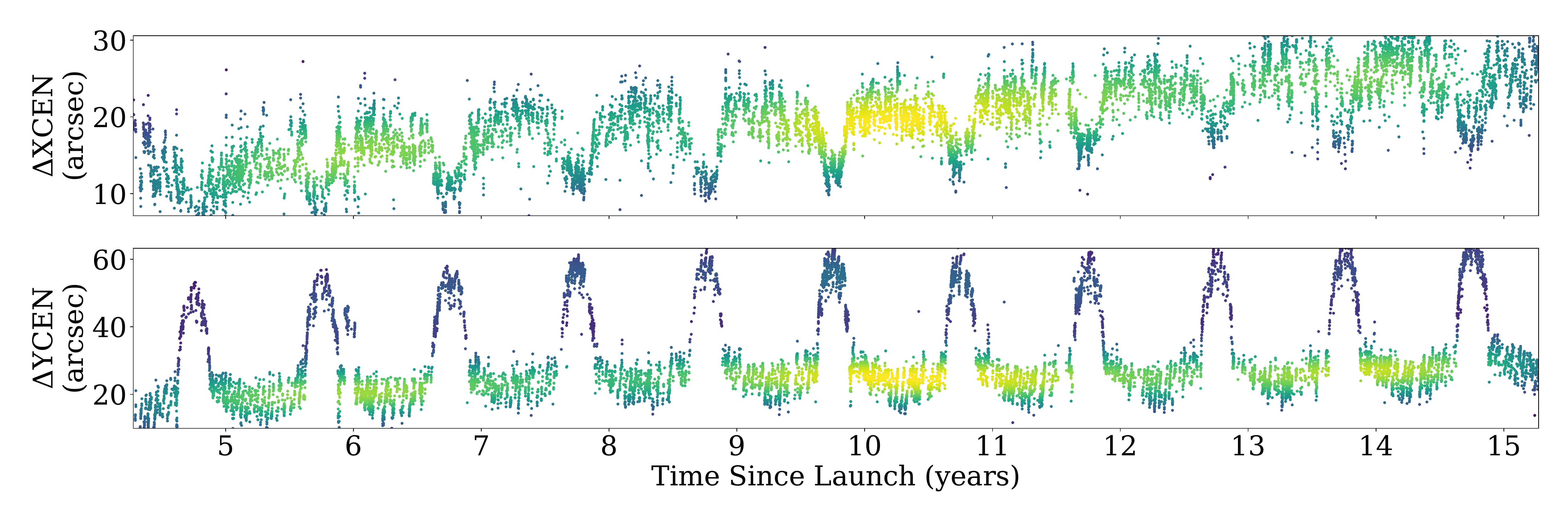}
\caption{{\bf Pointing updates as a function of time since launch (22 September 2006).} We plot \dxcen and \dycen as scatterplots (showing the middle 98\% of the data), coloring points by their density using a kernel density estimate. There are substantial secular and cyclic trends. This plots uses improved estimates of \xcen and \ycen calculated via the {\tt X\_COORDINATE} and {\tt Y\_COORDINATE} fields to update some incorrect (but easily fixed) values reported in headers. Not doing this correction would mix a large easily-fixed error involving a small amount of the data with the few-dozen arc-second more difficult-to-remove trend impacting most of the data.
The effects of these substantially incorrect \xcen and \ycen header values are shown in Figure~\ref{fig:application_update_pointing}.
Density colormap: Min \includegraphics[width=20pt,height=6pt]{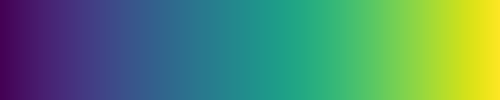} Max}
\label{fig:history}
\end{figure*}

\hinode pointing information is known to be inaccurate, as described in its data analysis guide~\cite[Appendix C]{hinodeDataAnalysisGuide}. However, the transformations fit between \hmi and \hinode let us provide updated relative pointing information. This pointing information is relative to \hmi, and is therefore limited by the accuracy of \hmi's pointing information, any optical distortions~\citep{Schou2012}, and the accuracy of our procedure. However, \hmi's observations include the limb of the Sun, which permit a more accurate determination of its pointing. We now explore updating \hinode's pointing information using the full 2011-2021 dataset.

\begin{figure}[t!]
    \centering
    \includegraphics[width=0.97\linewidth]{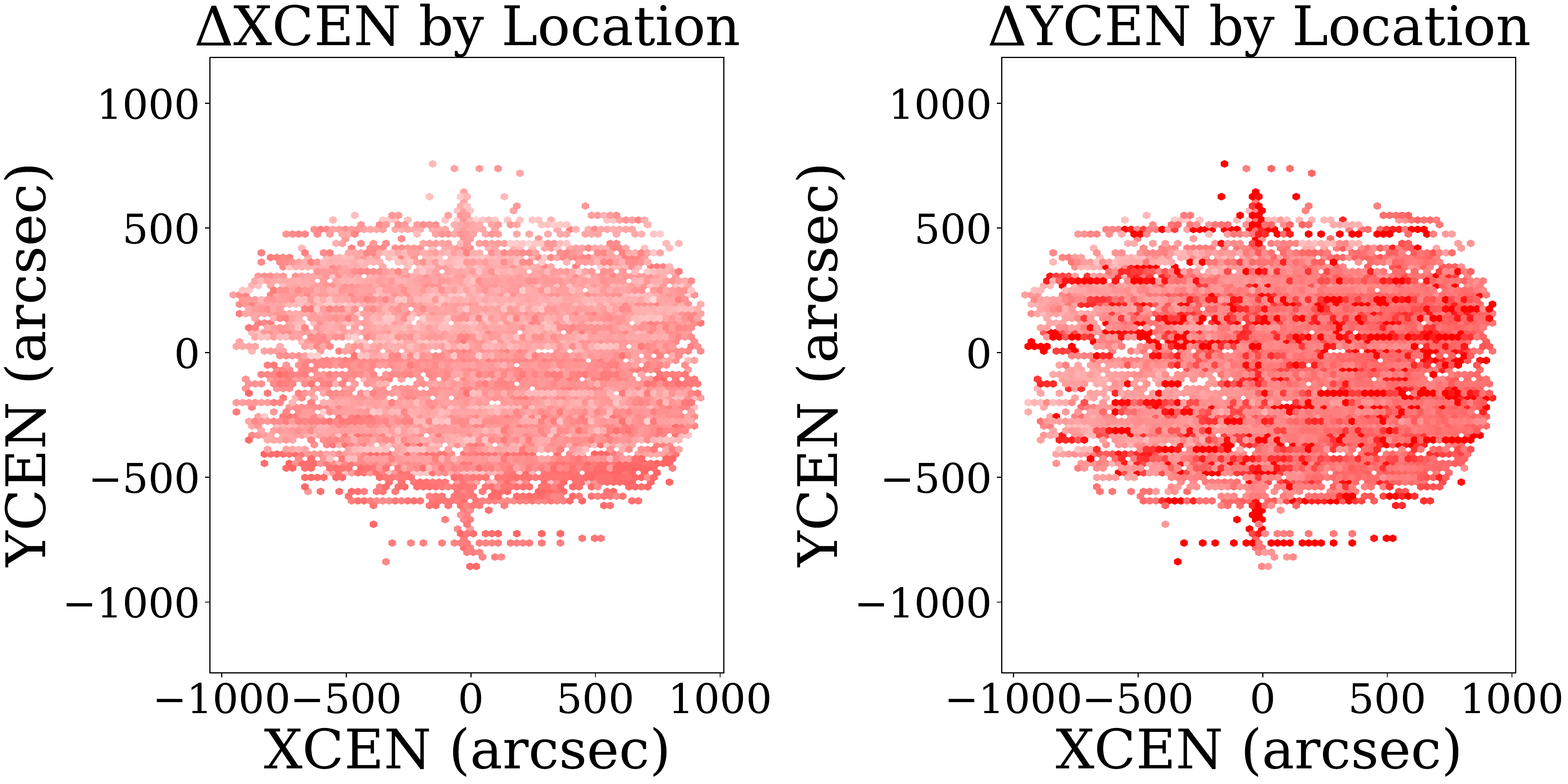}
    \caption{{\bf Plots of \dxcen and \dycen as a function of {\tt XCEN} and {\tt YCEN} locations.} For each location, we plot the averaged \dxcen (left), \dycen (right) per location. Colormap: -50\arcsec~ \includegraphics[width=20pt,height=6pt]{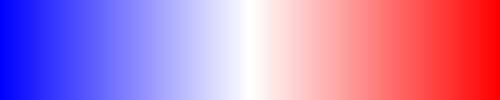}~50\arcsec . There is a strong positive bias, as indicated by the uniformly red color.}
    \label{fig:scatdxdy}
\end{figure}

\begin{figure*}[t]
\centering
\begin{tabular}{c}
\includegraphics[width=\linewidth]{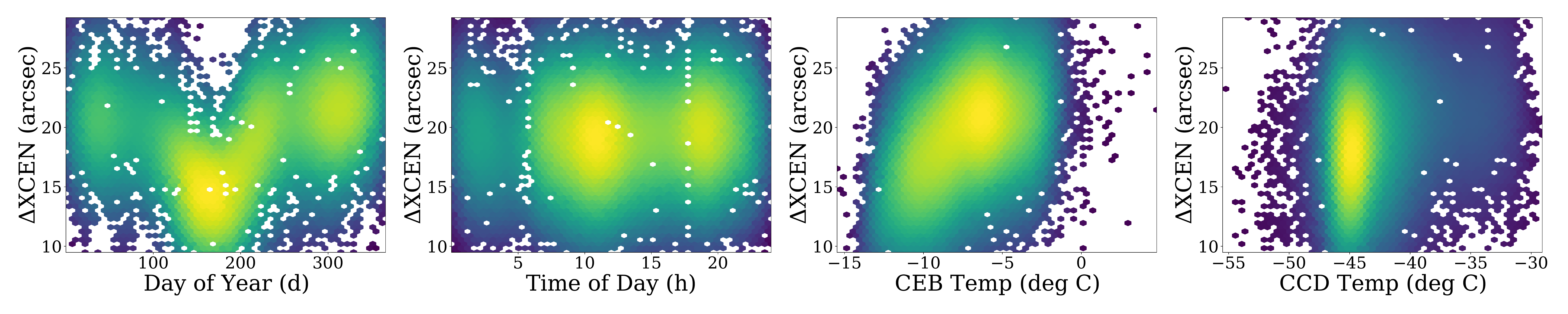} \\
\includegraphics[width=\linewidth]{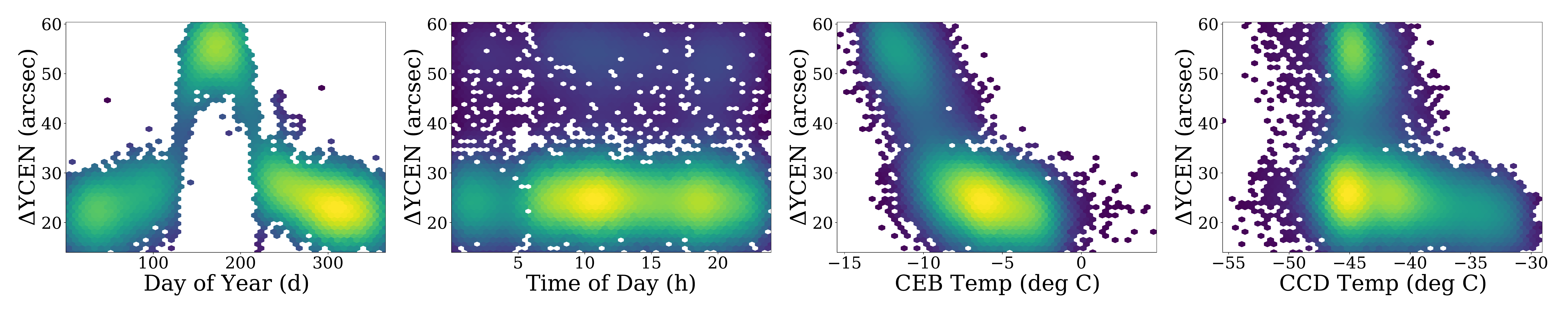} \\
\end{tabular}
\caption{{\bf Plots of the fit \dxcen (top) and \dycen (bottom) pointing residuals against multiple covariates.} The colors of the points show the density in the area, estimated with a kernel density estimate. We plot the residuals against day of year and time of day, as well as temperatures at the Camera Electronics Box (CEB) and Charge-coupled device sensor (CCD). Both $X$ and $Y$ residuals have a large shift during eclipse season mid-way through the year. This shift is well explained by change in temperature, suggesting that thermal expansion plays a role. To avoid distorting the plot with clearly-fixable errors and a small number of clear outliers, the plots use \xcen and \ycen re-calculated from the coordinate fields, rather than the header and show the middle 98\% of the data. Density colormap: White: Not observed; min \includegraphics[width=20pt,height=6pt]{viridis.png} max.
}
\label{fig:pointingtrends}
\end{figure*}

Just as one can warp \hinode data to the \hmi grid to enable applications such as the synthetic inversion system presented in \citep{Higgins2022}, one can also warp a grid of coordinates. The reported coordinates are provided in the Level 2 \hinode data products as {\tt X\_Coordinate} and {\tt Y\_Coordinate} as well as implicitly via the values {\tt XCEN} and {\tt YCEN}.  To update the coordinates, we warp the helioprojective coordinates from \hmi, which we are calculated via SunPy \citep{sunpy_community2020}. The transformation that was fit between \hmi and \hinode references the slit position as opposed to the scan index; however, recall that this can be transformed to the Level 2 \hinode data by dropping slit positions that were not observed. Our procedure does not account for relative spacecraft position, but we estimate that the relative position leads to an error of under $1\arcsec$. 

We illustrate this warping in Figure~\ref{fig:pointing}, showing alignment
for flux density ($\alpha B$), continuum intensity, and disk masks. The \hinode data is warped to the \hmi grid according to the fit transformation using a cubic spline for interpolation. The \hinode field strength is calculated by the product of intrinsic field strength {\tt Field\_Strength} and fill factor {\tt Stray\_Light\_Fill\_Factor}. A transformation that aligns both instruments' field strength (or similarly, their continuum images) results in substantial misalignment of the reported \hinode information but good alignment with our corrections (by construction).

Our updated coordinates can then be used to update pointing information {\tt XCEN} and {\tt YCEN}. We note, however, that a few scans have substantially incorrect data for {\tt XCEN} and {\tt YCEN} in the header but more accurate x and y coordinates in the {\tt X\_COORDINATE} and {\tt Y\_COORDINATE} fields; we thus recalculate {\tt XCEN} and {\tt YCEN} by the average of the x coordinates in the first and last columns, and similarly for the rows and the y coordinate. We calculate the updated values of {\tt XCEN} and {\tt YCEN} identically using our updated coordinates that are warped from \hmi. This usually does not change coordinates, except in the small number of cases with substantial mistakes.

We compute the residuals (i.e., $\Delta${\tt XCEN} = {\tt XCEN}$_\textrm{new}$ - {\tt XCEN}$_\textrm{reported}$ and similarly for \dycen). We plot the residuals as a function of mission time in Figure~\ref{fig:history} and disk location in Figure~\ref{fig:scatdxdy} (displayed as a hexbin plot, averaging within hexagonal bins, due to the large number of data points). The plot of (\dxcen, \dycen) as a function of time reveal strong annual oscillations in pointing on the order of a 30\arcsec~as well as a slow secular drift across over time. When plotted across the disk, these residuals are relatively consistent. Across the the dataset, the mean norm of the residual is $20.4\arcsec$~in x axis and $32.5\arcsec$~in the y axis. Our updated X and Y coordinates are near perfectly aligned with \hmi by construction (although they are dependent on the accuracy of our mapping).

\subsection{Exploring The Pointing Residuals}

Given the nature of the residuals shown in Figure~\ref{fig:history}, we then examine the pointing residuals \dxcen, \dycen in more detail. We initially explored temporal trends such as total of time since launch (which captures secular trends) as seen in Figure~\ref{fig:history} as well as time of year (which captures Earth-orbit effects) and time of day (which captures instrument orbit effects). We show \dxcen and \dycen plotted against the covariates in Figure~\ref{fig:pointingtrends}. Both residuals shift substantially during the middle of the year during the {\it Hinode} eclipse season. This shift is especially pronounced for the Y residual, but is also visible in the X as well. We note that this shift does {\it not} occur during \hmi's biennial eclipse season. After spotting this relationship, we also plotted the residuals against temperature data reported at the camera electronics box (CEB) and CCD, which are given by {\tt T\_SPCEB} and {\tt T\_SPCCD} in the Level 1 headers. There is a strong relationship between the residuals and temperature, especially at the CEB.

Similar relationships were observed by \cite{Mariska2016} when performing co-alignment between {\it Hinode}/EIS and {\it SDO}/AIA, who also pointed to thermal issues. Like our results, the range reported in \dxcen and \dycen were around 30\arcsec and 60\arcsec~respectively as reported by \cite{hinode2019achievements}. {\it SDO}/AIA and \hmi do not share optical pathways, nor do {\it Hinode}/EIS and {\it Hinode}/SOT-SP. Moreover, the two instruments onboard {\it Hinode} do not have the same pixel size. The close agreement in the size of the effect despite different optical pathways and pixel sizes suggests that the thermally-driven variance in pointing originates in {\it Hinode}'s AOCS. We discuss this further in Section~\ref{sec:discussion}.

\subsection{Explaining The Pointing Residuals}
\label{sec:explainingresiduals}
To more systematically explore the approach, we fit a linear model that takes covariates capturing time and temperature, and links them to the pointing residual (i.e., \dxcen and \dycen). We use total time, time of year, time of day, and the CEB and CCD temperature. Each covariate was independently standardized (i.e., to have a mean of zero and standard deviation of unity). This prevents any regularization from being disproportionately applied to covariates with larger intrinsic ranges of values. The fit coefficients can explain the relative role of the parameters.

The relationship between many covariates and the pointing residuals is nonlinear, and so we performed a basis expansion for each covariate. This includes a quadratic term as well as well as radial basis function that that softly bins the covariate into a set of discrete categories across the typical range of values. Each radial basis function covariate is the probability of the value according to a set of Gaussians evenly spaced from $-$2 to 2. Thus, for a single covariate, the final result is the concatenation of the original covariate, its square, and the radial basis functions, or 
\begin{equation}
\phi(x) = [x, x^2, p(x;\mu_1,\sigma), \ldots, p(x;\mu_{20},\sigma)]
\end{equation} 
where $\mu_i$ is the ith Gaussian's mean, $\sigma$ is an empirical parameter (set to 0.2), and $p(x;\mu,\sigma)$ is the Gaussian density.

Given the covariates $\{\xB_i\}_{i=1}^N$ and a corresponding residual $\{y_i\}_{i=1}^N$ for either the $x$ or the $y$ axis, we solve for a linear model with an $\ell_1$-regularized LASSO objective, 
\begin{equation}
\argmin_{\wB \in \mathbb{R}^F} \lambda ||\wB||_1 + \sum_{i=1}^N (y_i - \wB^T \Phi(\xB_i))^2
\end{equation}
where $\Phi$ applies $\phi$ to each coordinate of $\xB$ and concatenates, producing an $F$-dimensional vector. The $\ell_1$ penalty $\lambda ||\wB||_1$ on the fit promotes simplicity in the form of sparsity and ensures that only a few covariates are selected. Rather than pick the trade-off parameter $\lambda$ to maximize performance, we simply lower it until at most seven parameters are fit (picked empirically). We found that a few outliers (e.g., with pointing updates of 377\arcsec) led to substantially worse fits due to the outliers dominating the fit. We therefore only fit on data points where {\it both} \dxcen and \dycen were within the middle 99\% of the data ($7.5\arcsec \le$ \dxcen $\le 64.2\arcsec$ and $5.9\arcsec \le$ \dycen $\le 31.3\arcsec$). This removes of 206 out of the 12062 successfully registered scans. 

The residuals are relatively well-explained by a subset of temporal  and temperature-related variables. We report coefficients normalized so that the largest has value $\pm 1$. The Y residual is explained primarily by radial basis functions at mid-year, as well as a secular trend term. The fit model consists of: Time of Year radial basis function at day 180 (weight 1), another Time of Year radial basis function at day 158 (weight 0.78),  CEB Temperature (weight $-$0.32), and Time Since Launch  (weight 0.17), the squared CEB Temperature (weight 0.13) followed by squared covariates for Time of Year and Time since Launch, both with weights less than 0.05. The X residual is driven by a secular trend as well as a mid-year radial basis functions: the model is Time Since Launch (weight 1), a time of year radial basis function at day 158 (weight $-$0.83), then the CEB Temperature (weight 0.36), the square of the CEB Temperature (weight $-$0.11), with covariates for the time of year, squared time of day, and X scale getting weights of less than 0.05. We note that due to the strong correlation between CEB Temperature and time of year, the model may pick either. Once one variable is picked, the other one is not needed due to the $\ell_1$ regularization.

\begin{figure}
    \centering
    \includegraphics[width=\linewidth]{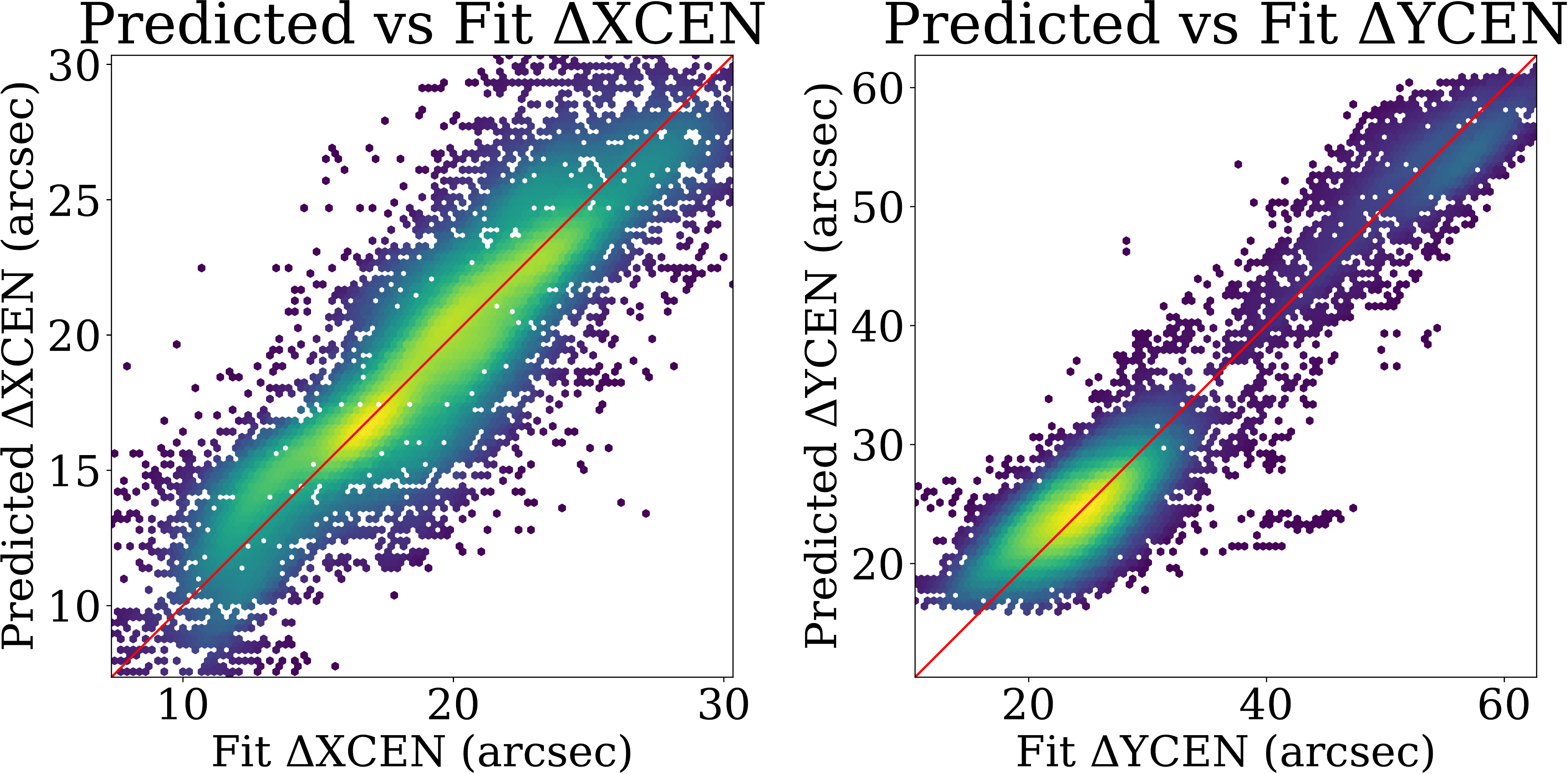} 
    \caption{{\bf Pointing update prediction performance.} We plot the cross-validated predictions of \dxcen, \dycen against the values fit by alignment. This plot shows the density of the region, estimated with a Kernel Density Estimate. The predicted values are generally quite accurate. Approximately 2\% of data are outliers and are not shown. Density colormap: Min \includegraphics[width=20pt,height=6pt]{viridis.png} Max.}
    \label{fig:predictedvsfit}
\end{figure}

\begin{figure}
    \centering
    \includegraphics[width=\linewidth]{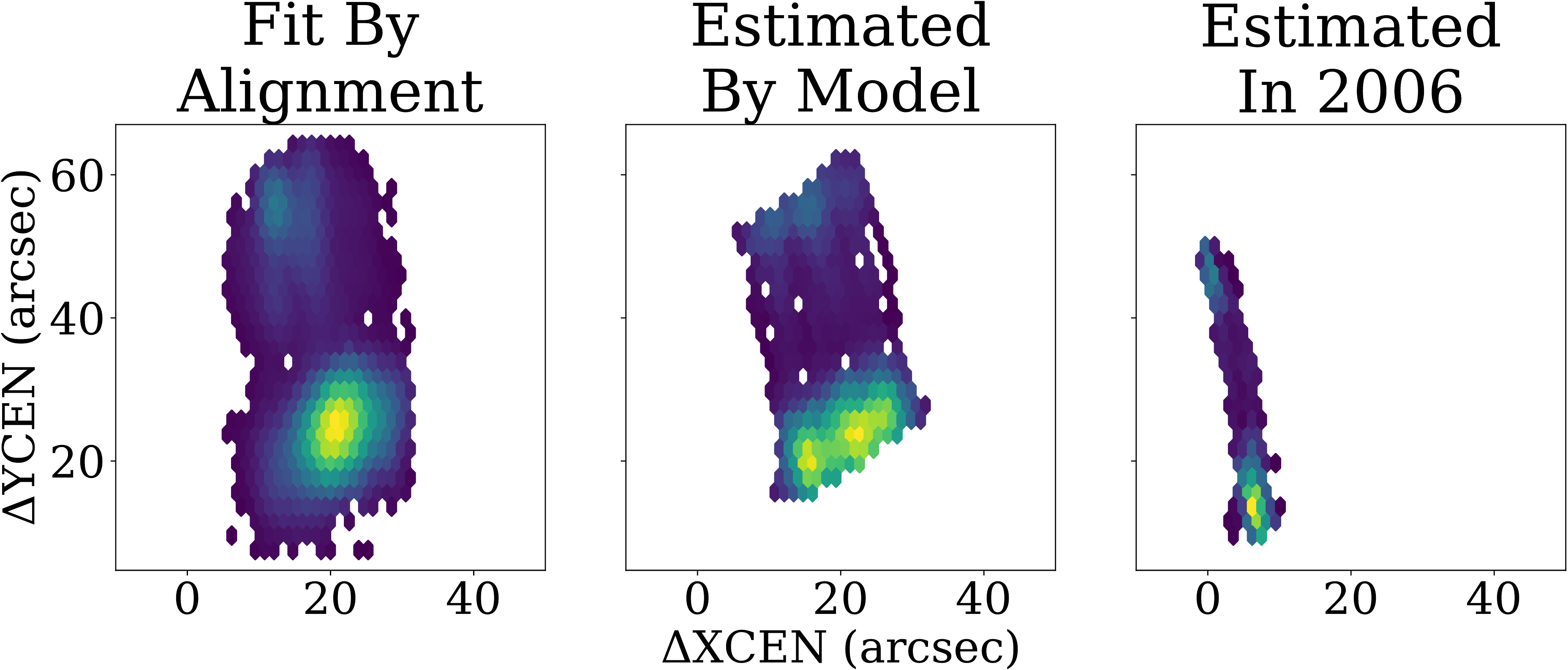} 
    \caption{{\bf Pointing Residuals.} (left) residuals fit via co-alignment; (middle) cross-validated predictions from temporal and thermal parameters; (right) cross-validated predictions on the same data, but with the data's timestamps shifted to approximately the time of launch. The predictions match the fit data fairly well but are pulled towards towards the mean (as is common minimizing the sum of squares). The extrapolations at time of launch shows a substantial reduction in residual. Since the predictions are cross-validated predictions (rather than the output of a single model), the time shift does not shift each datapoint identically. Approximately 2\% of original fit residuals data are outliers (e.g., \dxcen = 300\arcsec) and are not shown. Density colormap: Min \includegraphics[width=20pt,height=6pt]{viridis.png} Max.}
    \label{fig:rewind}
\end{figure}

\subsection{Predicting the Pointing Residuals}

Given importance of time of year and the total time and following \cite{Mariska2016}, we additionally fit a straightforward look-up table plus linear model consisting of two models. We refer to this as {\it Lookup-Plus-Linear}. The first model is a kernel regression (with $\sigma{=}3$ days) on the time of year, which functions like a smoothed look-up table to handle the bimodal behavior. The second model is an affine function (i.e., linear function plus bias) of the time since launch variable without any coefficient expansion. This linear function handles the secular trend found in the data. We fit the model stagewise: we make cross-validated kernel regression predictions to estimate the kernel regression models' output and then fit a linear model on the residuals between the target and the cross-validated predictions.

\begin{deluxetable}{ccccc}[t]
\caption{{\bf Mean absolute error (MAE) for prediction of the pointing residual.} We report 20-fold cross-validated MAEs for four models for producing the pointing residual: (zero) using the data as-is; (median) using the median correction from the training set; (LASSO) an $\ell_1$ model fit following \S\ref{sec:explainingresiduals}; (Lookup+Linear) an additive model combining an interpolated lookup-table on time-of-year and a linear model on the time since launch. While the initial data is substantially misaligned, a simple model can reduce the misalignment by an order of magnitude.}
\label{tab:residualpredict}
\tablehead{
\colhead{Scan} & \colhead{Zero} & \colhead{Median} & \colhead{LASSO} & \colhead{Lookup+Linear}}
\startdata
\dxcen & 19.3 & 4.4 & 2.1 & 2.1 \\
\dycen & 31.6 & 10.2 & 4.4 & 3.2 \\
\enddata
\end{deluxetable}

The LASSO and lookup-plus-linear models explain the data well, as shown by the cross-validated MAE for multiple models shown in Table~\ref{tab:residualpredict}. While directly co-registering existing \hinode scans to \hmi data will lead to better alignment, these results suggests that one can apply a deterministic correction to substantially improve pointing information. The current pointing information is off on average by about 30\arcsec~in Y. A constant correction reduces the error by $68\%$, and a simple model reduces the error by $90\%$. We show cross-validated predictions against the fit residuals in Figure~\ref{fig:predictedvsfit}, which shows reasonably tight agreement.

While \hmi was not available at the time of \hinode's launch, we can use our fit models to extrapolate pointing residuals back in time to the launch of {\it Hinode} in 2006. We additionally predict the pointing residual on data points that have been altered by subtracting whole years from the time since launch, so as to leave other variables (e.g., time of year) unaltered. We plot these results in Figure~\ref{fig:rewind}, showing results also with the Lookup-Plus-Linear model. The data shows substantial reduction in both \dxcen and \dycen, suggesting that there has been substantial drift over time. Nonetheless, it still shows the variation in \dycen that correlates with the {\it Hinode} eclipse season. 

\begin{figure}
    \centering
    \begin{tabular}{cc}
    \includegraphics[width=0.45\linewidth]{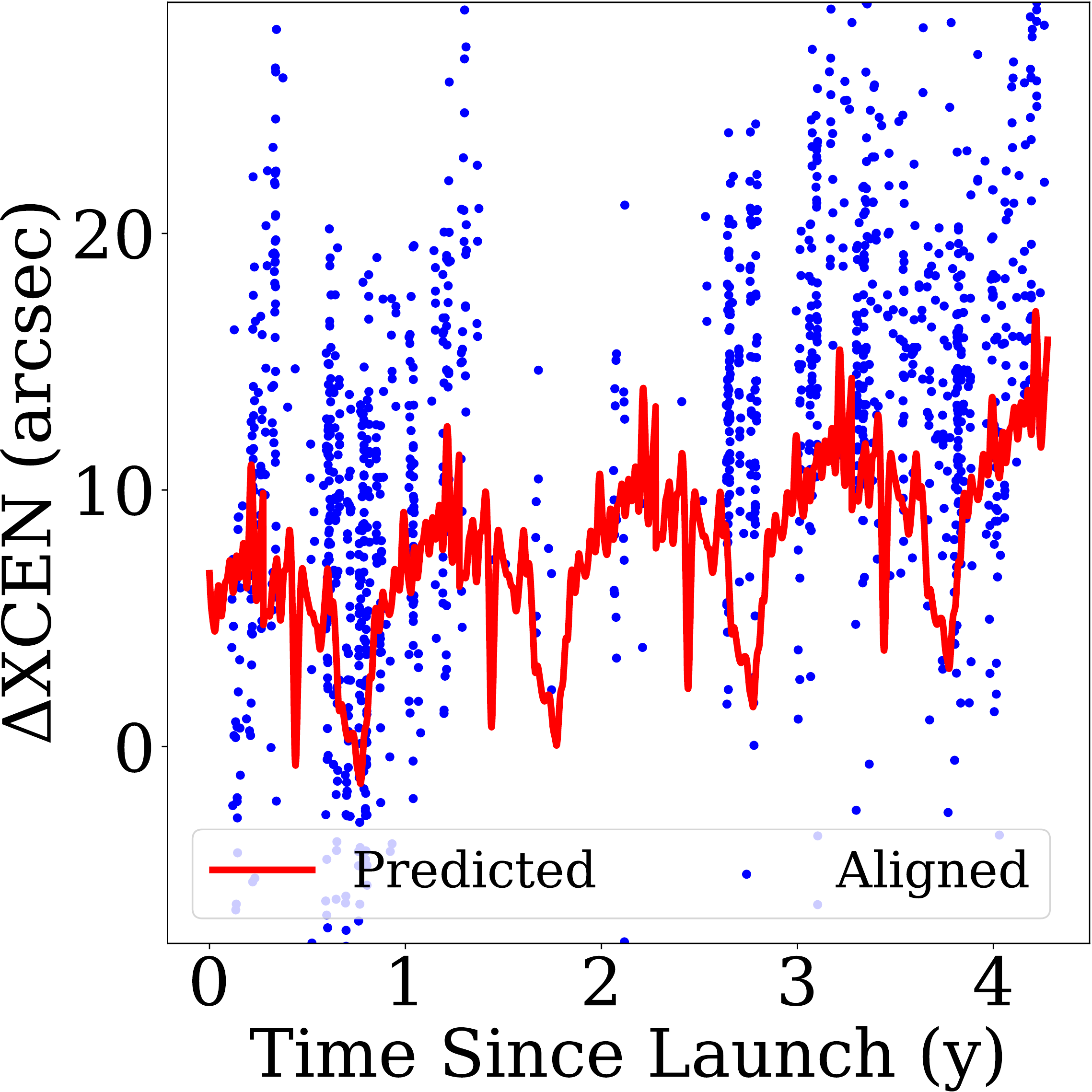} & 
    \includegraphics[width=0.45\linewidth]{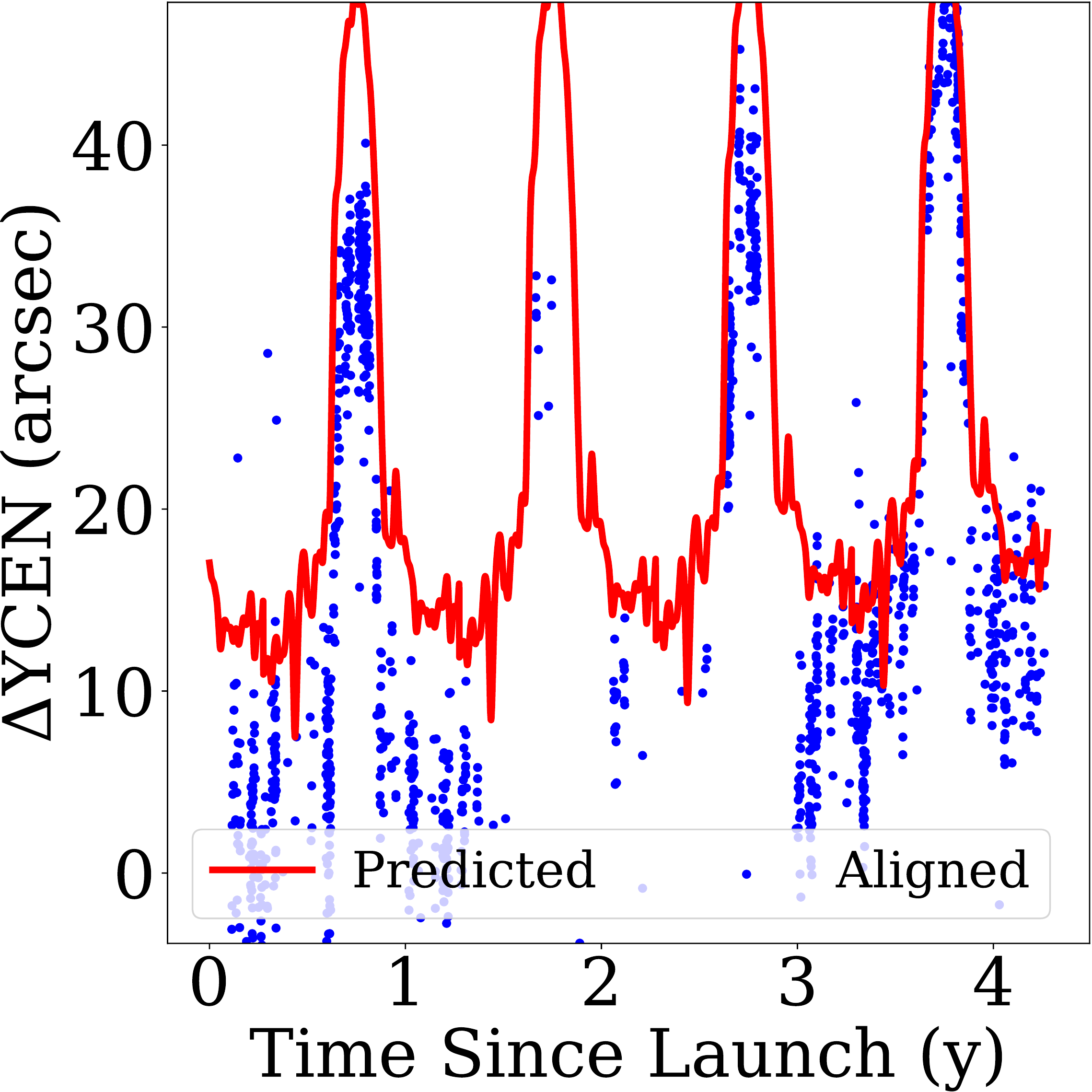}
    \\
    \end{tabular}
    \caption{
    {\bf Co-alignment of \hinode with \mdi.}
    We show a scatterplot of updated pointing information obtained from co-alignment with \mdi (in {\bf \textcolor{blue}{blue dots}}) alongside a prediction of the Lookup-Plus-Linear approach fit to the whole \hmi dataset (as a {\bf \textcolor{red}{red line}}). We plot both against time since launch (22 September 2006). As predicted by Figure~\ref{fig:rewind}, the pointing error via alignment gets substantially closer to zero near the launch date and continues to show the trends in Figure~\ref{fig:history}. The predicted pointing errors are reasonably accurate considering that this is an extrapolation.}
    \label{fig:pointing_mdi}
\end{figure}

For validation, we aligned \hinode data from 2006 -- 2010 with \mdi~\citep{scherrer1995solar}, using the MDI {\tt mdi.fd\_M\_96m\_lev182} data series. We took all of the corresponding \hinode data from 2006 -- 2010 meeting our criteria and used the procedure described in Section~\ref{sec:method} to co-align it with \mdi data. Since \mdi is a LOS magnetogram, it was necessary to change the quantities used in alignment. We matched the magnitude of the \mdi LOS field and with \hinode field strength and polarization. Additionally, due to the large change in scale between the two instruments, we found fewer inliers, and so to maximize the number of available correspondences, we used the full \hinode scan rather than the filtering by time. We plot the results in Figure~\ref{fig:pointing_mdi}. Among scans within a year of launch, the average \dxcen is 1.8\arcsec, and the average \dycen is 17.3\arcsec. The larger \dycen is driven by the eclipse-season jump: the averge \dycen in the first six months is just 1.65\arcsec.
The alignments that are fit to \mdi continue to show the trends of Figure~\ref{fig:history}, and the prediction of the model fit later in the mission produces a reasonable extrapolation to the  mission launch time.

\section{Physically-Relevant Implications}
\label{sec:implications}

\begin{figure}
    \centering
    \includegraphics[width=\linewidth]{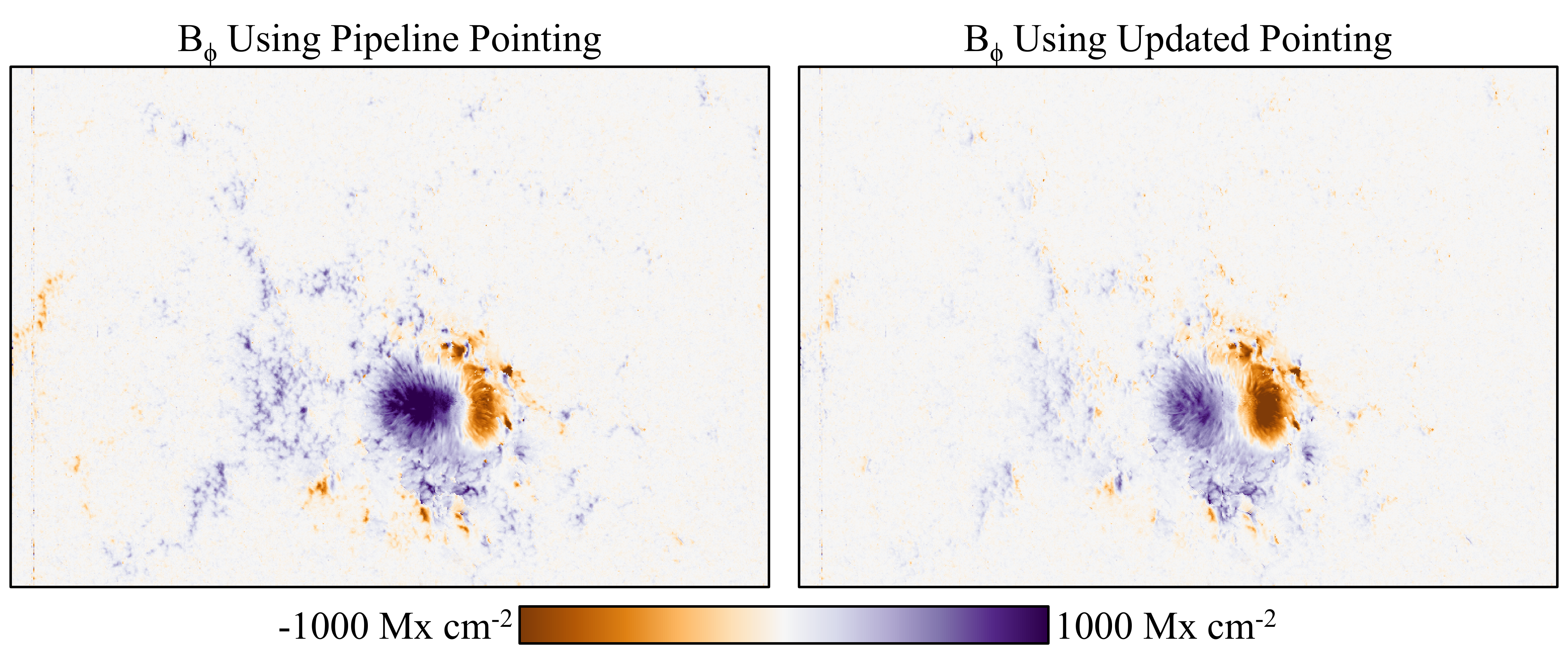}
    \caption{{\bf Impact of substantially incorrect {\tt XCEN} in the headers.} A small fraction of the \hinode scans have  pointing \change{in the header} that is substantially incorrect (by hundreds of arcseconds). Our previous figures have reported results with an upgraded \xcen calculated from the coordinates. One scan with an incorrect header appears on 2019 May 07, 06:49:34 TAI (the same AR, but not same scan as shown in Figure~\ref{fig:application_update_sample}). We plot $B_\phi$ calculated with both the pipeline header pointing~({\tt XCEN}~=~-339.6\arcsec) and our updated pointing~({\tt XCEN}~=~-586.7\arcsec). The substantial change dramatically changes the configuration. For reference, a more typical pointing update occurs a few hours later, at 08:39:32. In this more typical update, the pipeline pointing is~{\tt XCEN}~=~614.7\arcsec and our updated pointing is~{\tt XCEN}~=~589.1\arcsec.
    \label{fig:application_update_pointing}}
\end{figure}

\begin{figure*}
\centering
\begin{tabular}{cccc}
\includegraphics[height=1.7in]{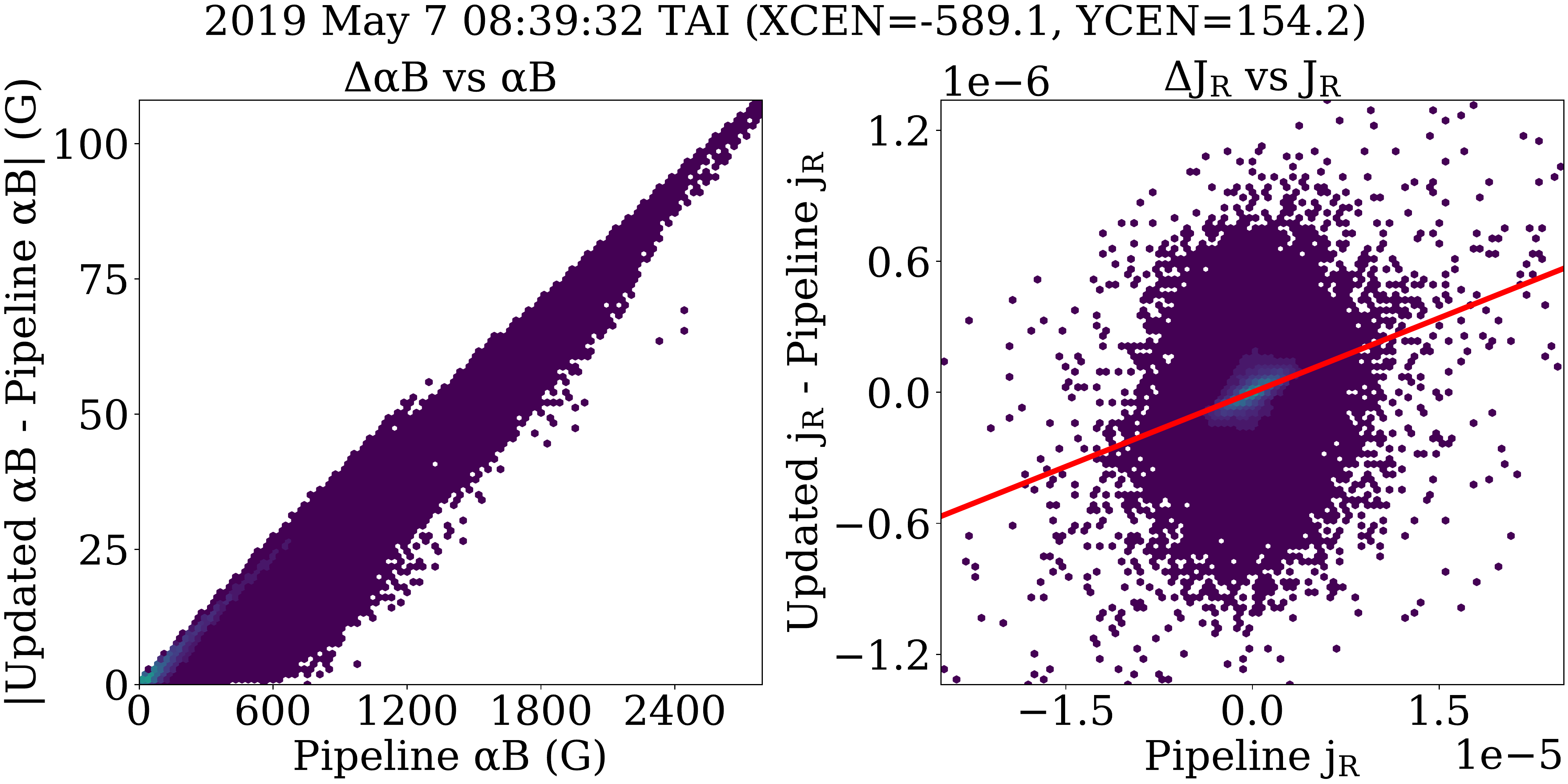} &
\includegraphics[height=1.7in]{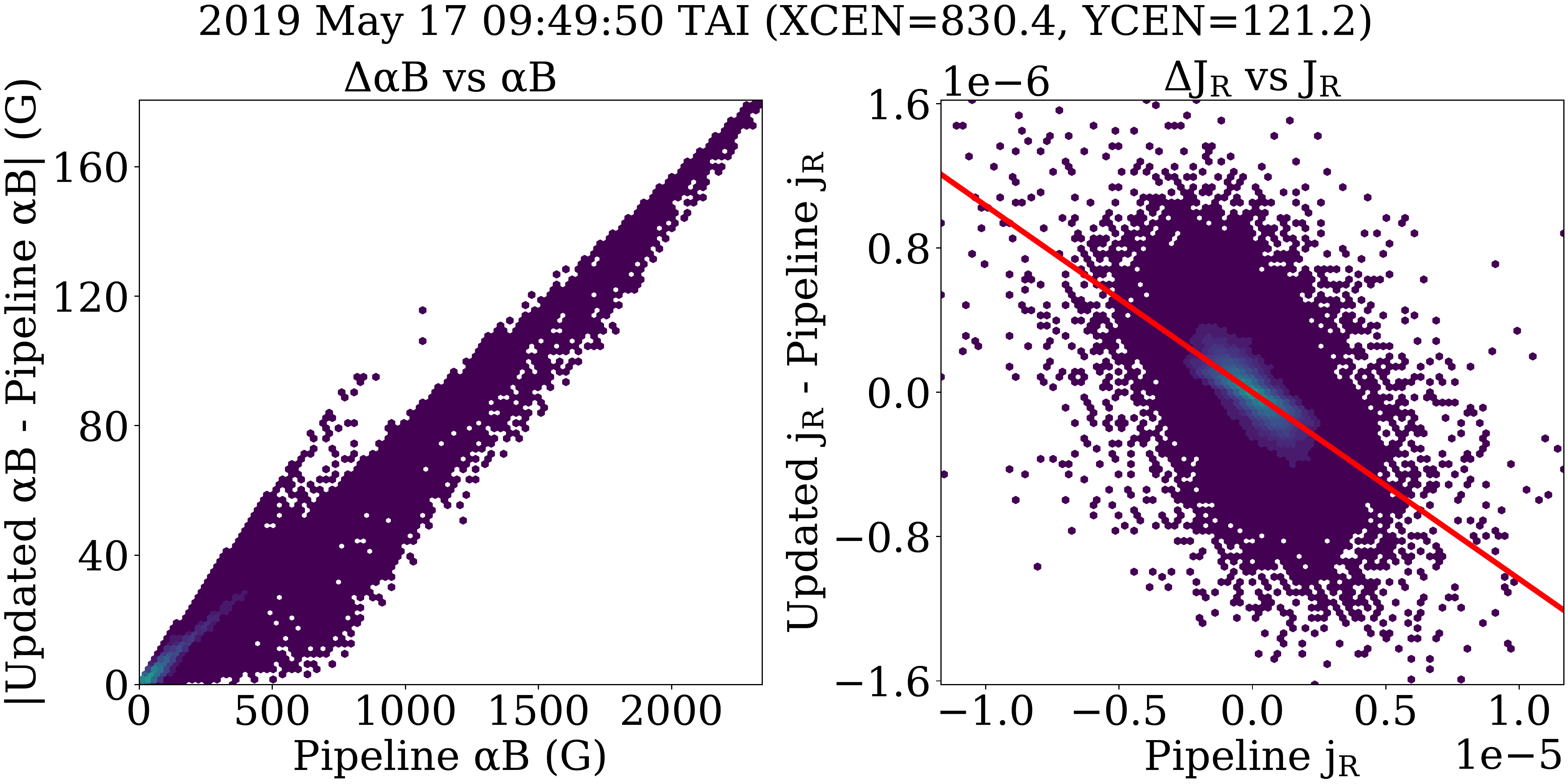}
\end{tabular}
\caption{{\bf Analysis of how updating the pointing information changes physical quantities}. We analyze two \hinode scans from early May 2019 on opposite sides of the Sun's disk (left:  2019 May 07 08:39:32 TAI at {\tt XCEN}=-589.1; right: 2019 May 17 09:49:50 TAI at {\tt XCEN}=830.4) with {\tt XCEN} determined via alignment. In each case, we plot the magnitude of the vector change in heliographic components against the magnitude on left and the change in radial current density against radial current density on right (along with a line of best-fit in \textcolor{red}{\bf red}).  Updating the pointing information changes the vector moderately (${\approx}5\%$) and the radial current density more (${\approx}10\%$). Moreover, the update to the radial current density is not consistent as seen by the skew in $\Delta j_R$-vs-$j_R$. On the the left panel's case, negative values of $j_R$ are changed to be more negative and positive values are changed to be more positive. On the right panel's case, this trend is reversed: negative values are made more positive positive and positive values are made less positive.  This trend holds true over larger amounts of data. Density legend: white: not observed; min \includegraphics[width=20pt,height=6pt]{viridis.png} max.
}
\label{fig:application_update_sample}
\end{figure*}

The corrections to the keywords {\tt XCEN}, {\tt YCEN} are typically on the order of 30 arcsec. The impact of this is different depending on where on the disk \hinode was observing. Near disk center, a shift of 30 arcsec in {\tt XCEN}, for example, corresponds to a change in longitude of less than $2^\circ$, while the same change in XCEN for an observation nominally at a longitude of $70^\circ$ is ${\sim}6^\circ$, and increases further approaching the limb. 

We note that some reported {\tt XCEN}, {\tt YCEN} are wrong due to suboptimal in-pipeline calculations. These can dramatically change the interpretation of regions, as shown for $B_\phi$ in Figure~\ref{fig:application_update_pointing}. In our analysis of the updates to the pointing in Section~\ref{sec:pointing}, we ignored these by re-calculating {\tt XCEN}, and in this section, we exclude such data from our analysis to avoid conflating the dozen-of-arcsecond changes impacting most of the data with the hundreds-of-arcsecond changes impacting some of the data that can be fixed by a quick re-calculation.

\begin{figure*}
\centering
\begin{tabular}{ccc}
\includegraphics[height=1.95in]{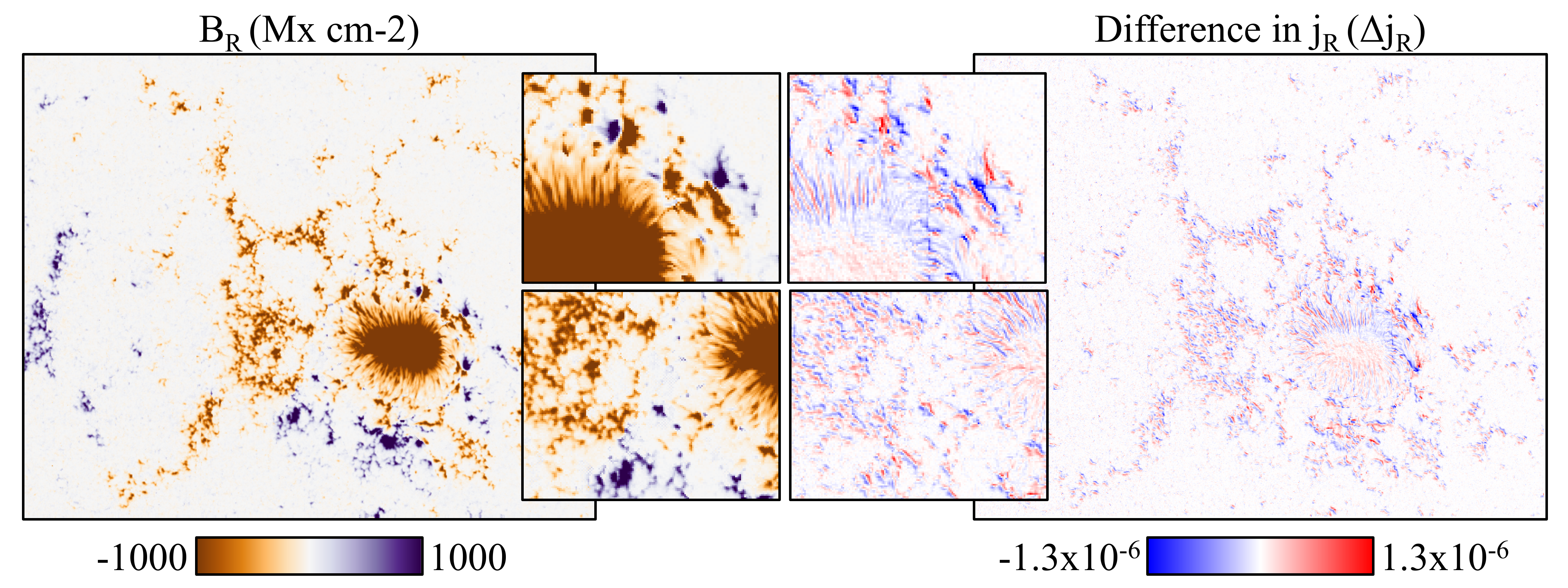} & ~~ &
\includegraphics[height=1.95in]{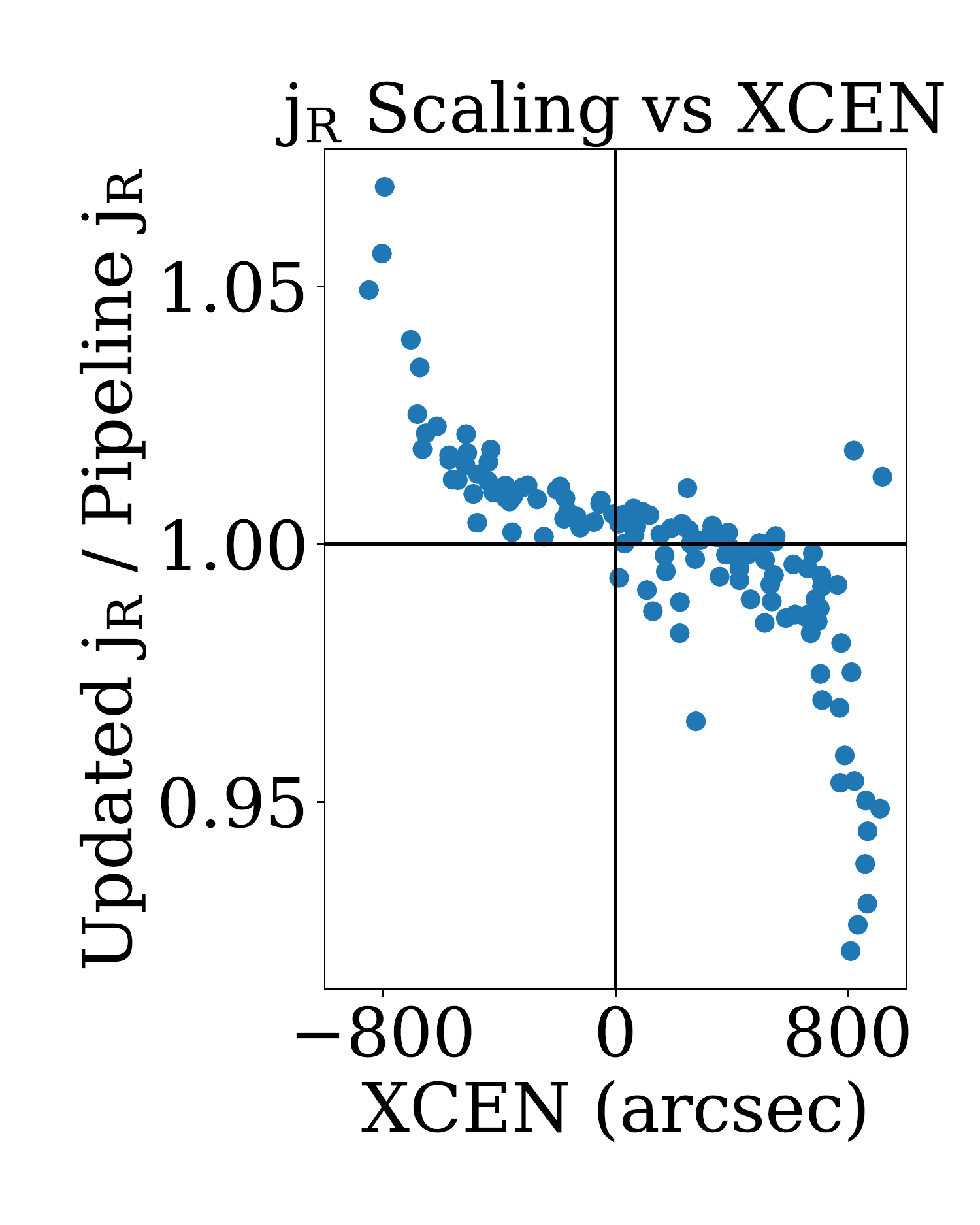}
\end{tabular}
\caption{{\bf Illustration of the non-uniform changes to $j_R$}. {\it Left two panels}: we plot $B_R$ for context and the difference in $j_R$, along with zoom-ins in between the images. The changes in $j_R$ are {\it not} spatially uniform, but exhibit distinct spatial trends consisting of ridges. {\it Right panel:} The overall relative scaling between $j_R$ calculated with the updated pointing information compared to $j_R$ calculated with the pipeline pointing information. On the left hemisphere, the new $j_R$ is almost always bigger; on the right, it is almost always smaller.}
\label{fig:application_update_spatial} 
\end{figure*}

Changes even at the scale of dozens of arcseconds, have physical implications, especially closer to the limb. We analyzed two active regions acquired on 2019 May 7 and 2019 May 17. Given the updated coordinates, one can re-compute the heliographic components of the field as well as other quantities such as the radial current density ($j_R$), which is important for determining the free energy in the corona. We plot the magnitude of the vector difference in $\alpha \BB$ (i.e., $||[\alpha B_\phi-\alpha B'_\phi, \alpha B_\theta-\alpha B'_\theta,\alpha B_R - \alpha B'_R]||$) against the magnitude $||\BB||$, as well as the signed change in $j_R$ (i.e., $j_R' - j_R$) against $j_R$. We show density plots in Figure~\ref{fig:application_update_sample} for each active region. The vector difference has a magnitude of ${\approx}5\%$ while the radial current density changes by ${\approx}10\%$. This difference occurs because, in addition to the heliographic components of the field, the distance between the pixels is changed. The change in inter-pixel distance alters the finite differences used to determine the radial current density. The change in radial current density, moreover, is not uniformly spread across the scan but instead appears as spatially distinct patterns as shown in Figure~\ref{fig:application_update_spatial}(left).

The positive bias of the corrections leads to bias in estimates of radial current density. On one limb, the updated inter-pixel distances always become smaller and on the other limb, they always become larger, resulting in the skew shown in Figure~\ref{fig:application_update_sample}(right). We investigate this further with a set of 126 random scans from the dataset that: (a) have $-200\arcsec <$ {\tt YCEN} $<200\arcsec$, to avoid {\tt YCEN} to be a confounding factor; (b) have \dxcen $> 10$ so as to avoid small changes (although most scans satisfy this); (c) do not have {\it enormous} changes in pointing (\dxcen $> 100\arcsec$ or \dycen $> 100\arcsec$) that are due to a mis-coding of {\tt XCEN} from the pipeline pointing. For each, we re-compute the radial current density $j'_R$ and compare it with the radial current density $j_R$ as computed with the pipeline pointing. We then compute the optimal linear scaling from the pipeline to updated quantities (including an offset term), and plot this scaling as a function of disk location. As can be seen in Figure~\ref{fig:application_update_spatial}(right), there is a strong correlation between disk location and the scaling. This means that regions observed close to the west limb will have a total current which is underestimated when using the original pointing information, while those observed close to the east limb will have the total current overestimated.

We present here an illustration only of how recomputing the heliographic components of the field with the updated pointing changes the results, but the heliographic components of the field rely on a disambiguation algorithm that uses the radial component of the current density to determine the most likely solution.  Thus the disambiguation can also be impacted by the change in the pointing, leading to even large changes in the heliographic components than shown here.

\section{Discussion and Conclusions}
\label{sec:discussion}

This paper has presented an approach for aligning and thus spatially calibrating Level 2 \hinode data and \hmi data. This is challenging, not only because \hinode is a scanning slit spectrograph but also because the instruments have varying characteristics. Indeed, as \cite{dalda2017statistical} and \cite{Higgins2022} point out, the magnetograms produced by the pipelines of the two instruments used in this study differ not only in spatial resolution but also key implementation details. These differences result in systematic differences in structures such as plage~\citep{abgm_etal_2021,btrans_bias_2}.

Nonetheless, the approach presented does well at alignment and enables backing out both scan-invariant parameters in the form of relative scaling as well as scan-variant parameters in the form of pointing and rotation. The method's success is built on well-understood techniques from computer vision of correspondence and geometric model fitting that generalize well and have formed the backbone of many of computer vision's pre-deep learning successes. We do stress, though, that proper accounting for \hinode's scanning slit mechanism is critical for the success of this method, and that while the \hinode Level 2 data is presented in an image-like format, it cannot be directly treated as such for alignment. Earlier versions of this study empirically found noticeably worse performance when slit position was not accounted for. This need to properly understand instruments underscores the need for careful, sustained interdisciplinary collaboration in order to obtain correct results. Accounting for all the particulars of the instrument separates this works' alignment from our earlier work~\citep{Higgins2022} and reduces accurate alignment's dependence on post-processing steps such as optical flow. However, we do note that pixel-perfect alignment of all parts of the image (e.g., for the creation of synthetic magnetogram training) would require an additional optical flow step due to the non-uniform evolution and flow of the Sun. 

We note that the techniques applied in this paper may help in other challenging solar co-alignment settings. One appealing aspect of the technique presented is that correspondence extraction is robust to substantial differences between the data that are being aligned, as illustrated by the alignment of field strength from \hinode and flux density from \hmi (which are different quantities) and inclination from \hinode and \hmi (which are different in each instrument's data due to the treatment of fill factor, as noted by~\cite{dalda2017statistical}) and \cite{btrans_bias_2}. Moreover, as explored in this work, the use of point correspondences enables the seamless use of multiple modalities via the concatenation of correspondences as well as the handling of temporal evolution by the removal of correspondences between pixels taken at substantially different times. One benefit for large-scale analysis not demonstrated in this work is that the outlier-robustness makes this technique particularly effective even in the presence of large corrupted regions. In the setting for which these techniques achieved their first success (large-scale reconstruction~\cite{Snavely07,schoenberger2016sfm}), this robustness enabled them to reconstruct large-scale 3D models of landmarks in the presence of tourists, vehicles, and changing seasons.

If the \hmi scaling is correct, our experiments across thousands of scans spanning nearly a decade of the {\it Hinode} mission suggest that the \hinode scaling is smaller than the values reported in the Level 2 headers and close to the nominal Level 1 values. This result contradicts the results found by~\cite{centeno2009hinode}, who cross-calibrated \hinode with {\it Hinode}/SOT-FG BFI, which also observes a limited field of view (218\arcsec~by 109\arcsec) using the same base telescope. In the intervening time, the community has gained access to \hmi, which observes the full ${\approx}$2000\arcsec-wide solar disk, albeit from a different vantage point.

Our procedure can only provide information about {\it relative} scaling, and cannot conclusively answer questions about {\it absolute} scaling. However, \hmi has several advantages over {\it Hinode}/SOT-FG BFI and \hinode for serving as the source of reference measurements. First, \hmi's full disk observations show the limb of the Sun. This provides strong constraints on scaling compared to the narrower fields of view on active regions observed by {\it Hinode}/SOT. As a secondary consideration, \hmi's observations function far more like a global shutter (i.e., capturing a consistent snapshot of the Sun at a particular time): the Stokes vector observations used for HMI magnetograms are measured in approximately 2 minutes (although the data series we use averages multiple aligned observations over 12 minutes to reduce noise). In contrast, \hinode's use of a scanning slit means that rightmost pixels are acquired dozens of minutes later than the left ones. These complexities in the spatial dimension are necessary for \hinode's far higher spectral resolution and sampling, but they make \hmi a more compelling global reference instrument.

The translations we fit with our model provided updated estimates of the pointing information of \hinode for over twelve thousand scans spanning a decade. Currently, this pointing information is obtained by pointing obtained from other instruments aboard {\it Hinode}. As we showed, the errors in the current pointing information leads to incorrect estimates of physical quantities, with the particular example of radial current density. Our results suggest that the relative pointing calibration has both drifted slowly over time and that changes in temperature cause substantial seasonal changes in pointing around \hinode's eclipse season (and not during {\it SDO}'s eclipse seasons, which occur at different times of the year). These changes in pointing are well predicted by temperature onboard {\it Hinode}, suggesting that thermal expansion is the underlying cause.

Similar pointing corrections were reported for {\it Hinode}/EIS by \citep{Mariska2016,hinode2019achievements}, who obtained these corrections by cross-calibrating {\it SDO}/AIA using different methods. Our results show similar issues on \hinode with similar scales and help narrow down the cause. Given that \hinode and {\it Hinode}/EIS do not share optical pathways and have different pixel sizes, it seems less likely that internal calibration for the instruments is shifting identically with precisely the right size. Instead, it seems more likely there is a single shift in {\it Hinode}'s attitude and orbital control system.

These results underscore the value of having multiple instruments with varying capabilities: the pointing issue would be difficult to identify and explain without {\it SDO}. The co-observation of {\it SDO} further enable us to extract data: our updated estimates of scaling and pointing information for \hinode opens the door to a larger and more precisely aligned version of the joint dataset that powered the SynthIA system developed by \cite{Higgins2022}.

\par \noindent {\bf Acknowledgments:} 
This work is an interdisciplinary collaboration that is the result of the NASA Heliophysics DRIVE Science Center (SOLSTICE) at the University of Michigan under grant NASA 80NSSC20K0600. KDL and GB acknowledge support from
Lockheed Martin Space contract \#4103056734 for Solar-B FPP Phase E.

All \hmi data used are publicly available from the Joint Science Operations Center (JSOC) at Stanford University supported by NASA Contract NAS5-02139 (HMI), see \url{http://jsoc.stanford.edu/}.
{\it Hinode} is a Japanese mission developed and launched by ISAS/JAXA, with NAOJ as domestic partner and NASA and STFC (UK) as international partners. It is operated by these agencies in co-operation with ESA and NSC (Norway). Data and models used will be archived at the U-M Library Deep Blue data repository. All datasets will be given Digital Object Identifiers (DOIs).


\bibliography{main}{}
\bibliographystyle{aasjournal}



\end{document}